\documentclass[lettersize,journal]{IEEEtran}
\usepackage{amsmath,amsfonts}
\usepackage{algorithmic}
\usepackage{algorithm}
\usepackage{array}
\usepackage[caption=false,font=normalsize,labelfont=sf,textfont=sf]{subfig}
\usepackage{textcomp}
\usepackage{stfloats}
\usepackage{tabularx}
\usepackage{enumitem}
\usepackage{tikz}
\usetikzlibrary{arrows.meta,positioning,fit,calc,shapes,backgrounds}
\usepackage[table]{xcolor}
\usepackage{multirow}
\usepackage{graphicx}
\usepackage{threeparttable}
\usepackage{cite}
\usepackage{makecell}
\usepackage{amssymb}
\usepackage[colorlinks=true,linkcolor=blue,citecolor=blue,urlcolor=blue]{hyperref}
\usepackage{orcidlink}

\newcommand{\fullcircle}{%
  \tikz[baseline=-0.6ex]{
    \fill[black] (0,0) circle (3pt);
  }%
}

\newcommand{\halfcircle}{%
  \tikz[baseline=-0.6ex]{
    \begin{scope}
      \clip (0,0) circle (3pt);               
      \fill[black] (-3pt,-3pt) rectangle (0,3pt); 
    \end{scope}
    \draw[thick] (0,0) circle (3pt);          
  }%
}

\newcommand{\emptycircle}{%
  \tikz[baseline=-0.6ex]{
    \draw[thick] (0,0) circle (3pt);
  }%
}

\begin{document}

    \title{Cooperative Safety Intelligence over V2X Networks: A Survey}
  \author{Jiaxun~Zhang,~\protect\orcidlink{0009-0002-3234-2569}
      Qian~Xu,~\protect\orcidlink{0000-0002-3374-4108}
      Zhenning~Li,~\IEEEmembership{Member,~IEEE,}~\protect\orcidlink{0000-0002-0877-6829}
      Yuan~Wu,~\IEEEmembership{Senior Member,~IEEE,}~\protect\orcidlink{0000-0001-6661-9461}
      Chengzhong~Xu,~\IEEEmembership{Fellow,~IEEE,}~~\protect\orcidlink{0000-0001-9480-0356}
      Keqiang~Li~\protect\orcidlink{0000-0001-6223-5401}

    \thanks{Corresponding author. E-mail: \texttt{zhenningli@um.edu.mo}.

    Jiaxun Zhang, Qian Xu, Zhenning Li, Yuan Wu, and Chengzhong Xu are with the State Key Laboratory of Internet of Things for Smart City, University of Macau, Macau.
    Jiaxun Zhang, Qian Xu, and Zhenning Li are also with the Department of Civil and Environmental Engineering, Faculty of Engineering, University of Macau, Macau.
    Zhenning Li is also with the Department of Artificial Intelligence, Faculty of Information Science and Computing, University of Macau, Macau.
    Yuan Wu is also with the Department of Electronic and Communication Engineering, Faculty of Information Science and Computing, University of Macau, Macau.
    Chengzhong Xu is with the Department of Computer Science, Faculty of Information Science and Computing, and the Department of Electrical and Computer Engineering, Faculty of Engineering, University of Macau, Macau.
    Keqiang Li is with the School of Vehicle and Mobility, Tsinghua University, Beijing, China.}}
\maketitle

\begin{abstract}
Vehicle-to-Everything (V2X) cooperation is reshaping traffic safety from an ego-centric sensing problem into a networked intelligence problem involving distributed sensing, cooperative perception, and coordinated decision-making. This survey reviews recent progress in V2X-enabled cooperative safety intelligence through a unified Sensor–Perception–Decision (SPD) framework, which characterizes how distributed observations are exchanged, fused, calibrated, and transformed into safety-ready evidence for risk-aware intervention. Within this framework, V2X networking conditions, multi-modal sensing, cooperative perception, and decision-making are analyzed as coupled components of a safety-intelligence pipeline. Cross-layer communication and networking constraints, including latency, synchronization, bandwidth, packet or feature loss, reliability, trust, and PQoS, are explicitly considered to assess whether shared evidence remains timely, robust, and actionable. Compared with prior V2X safety surveys, this work organizes the literature around a formal SPD safety loop and synthesizes representative methods, datasets, benchmarks, and platforms under communication-aware constraints from 2017 to 2026. It further examines semantic and task-oriented evidence exchange, bandwidth-efficient cooperative perception, ISAC-enabled V2X systems, safety-ready outputs, and evaluation practices under practical conditions such as delayed or lossy messages, pose misalignment, NLOS coverage, and heterogeneous deployment. The survey concludes with a roadmap toward scalable data infrastructure, PQoS-aware evaluation, embodied predictive intelligence, and trustworthy human-in-the-loop cooperation for next-generation V2X safety systems.
\end{abstract}


\section{Introduction}
Traffic safety is fundamental to human dignity, social stability, and sustainable development, yet it remains a persistent challenge for Intelligent Transportation Systems (ITS), where automation and reliance on sensors and communication networks introduce new vulnerabilities. Under \textit{Vision Zero}, traffic safety is viewed as a shared responsibility among vehicles, roads, and people through systemic design~\cite{bendiab2023autonomous,fang2023vision}. This requirement becomes more critical in cooperative ITS, where Vehicle-to-Everything (V2X) enables traffic accident detection, anticipation, and beyond-visual-range safety services~\cite{zhang2025eyes,zhang2025latte,guan2025domain}, building on decades of V2X access technology and regulatory development~\cite{machardy2018v2x}. As safety shifts from post hoc mitigation to proactive prevention, autonomous vehicles need to move beyond reactive pipelines and reason under uncertainty~\cite{wang2020many}. {Accordingly, this survey focuses on application-level vehicle and traffic safety enabled by V2X, while treating communication reliability and trust as key enabling constraints~\cite{fang2023vision,soto2022survey}.}

Conventional perception-centric systems have made significant progress but remain constrained by their physical and informational boundaries~\cite{caillot2022survey,han2023collaborative}. Even with high-resolution cameras, radar, and LiDAR, the ego vehicle’s field of view is limited by occlusions caused by other vehicles, roadside structures, and complex urban layouts \cite{chu2025occlusion,liu2025towards}. These spatial restrictions shorten the anticipation horizon. Moreover, decisions based solely on local observations often lack the broader situational context needed for robustness: surrounding agents are difficult to predict, intent can be ambiguous, and sudden maneuvers may occur too close for safe intervention~\cite{ghorai2022state}. Consequently, many models—though competitive in benchmark accuracy—still deliver inconsistent real-world safety performance~\cite{chen2025qctf}. Effective forecasting benefits from providing foresight early enough to influence decisions and reduce risk, so that predictive insight can meaningfully contribute to crash avoidance \cite{xiao2023overcoming,fang2023vision}.

V2X offers a transformative opportunity to strengthen this predictive loop by turning safety from an individual vehicle attribute into a distributed system capability~\cite{min2025safe, hawlader2025poster}. Through vehicle-to-vehicle (V2V) messages and vehicle-to-infrastructure (V2I) feeds, connected systems exchange raw states and structured safety evidence, extending the perceptual horizon beyond line of sight and enabling an ego vehicle to anticipate hidden objects, braking events several cars ahead, or unsafe maneuvers in adjacent lanes~\cite{boban2018connected,ghorai2022state,fu2021survey}. {From a communication and networking perspective, such cooperative benefits depend on low-latency transmission, time synchronization, bandwidth allocation, packet reliability, and scalable multi-agent connectivity, which jointly determine whether shared evidence remains timely, credible, and usable for downstream perception and decision modules~\cite{noor20226g,shi2022sparse,alsudani2023wireless,gyawali2020challenges}. More recently, semantic and task-oriented V2X communications have further shifted the focus from data-centric transmission to context-relevant and task-useful safety evidence exchange, supporting scalable cooperative perception and coordinated responses under wireless access constraints that shape robustness and deployment feasibility~\cite{xu2023semantic,gimenez2024semantic,yusuf2024vehicle,clancy2024wireless}.}

Earlier studies have explored related paradigms such as \emph{cooperative perception}, \emph{vehicle–infrastructure cooperation}, and \emph{collaborative intelligence}. Cooperative perception frameworks~\cite{arnold2020cooperative, xu2024v2x} extend an ego vehicle’s visibility by sharing sensor features among vehicles and roadside units (RSUs), enabling awareness beyond line-of-sight (BLOS) occlusions. Vehicle–infrastructure cooperation~\cite{li2022v2x, chang2023bev} focuses on harmonizing vehicle trajectories and infrastructure sensing to achieve more consistent situational awareness and smoother control at intersections and highways. Collaborative intelligence, supported by edge and cloud computing~\cite{ lucas2022analytical, gharsallah2024mvx}, emphasizes distributed processing and AI inference across connected entities to improve efficiency and scalability in V2X networks. While these directions have enhanced connectivity and perception capabilities, they largely address separate layers of the system: cooperative perception optimizes sensing, vehicle–infrastructure cooperation improves coordination, and collaborative intelligence enhances computation~\cite{bejarbaneh2024exploring,yu2022review}. However, a unified notion of how sensing, prediction, and decision interact within safety-critical time budgets remains underdeveloped~\cite{fu2021survey, abdi2024advancing}.

\textbf{Cooperative Safety Intelligence} builds upon this foundation as a distributed safety ecosystem in which vehicles, infrastructure, and cloud resources use V2X communication to transform local observations into shared safety evidence and coordinated interventions. As used in this survey, it unifies sensing, perception, and decision-making across connected entities to support predictive, cooperative, and trustworthy operation toward \emph{Zero-Accident Mobility}. {Communication and networking serve as the enabling substrate of this ecosystem, determining what information can be shared, when it arrives, how reliable it is, and whether it can be effectively consumed by downstream perception and decision modules.} This perspective provides the organizing lens of the \textbf{Sensor--Perception--Decision (SPD)} framework, in which cross-cutting constraints---including {network latency, synchronization, bandwidth, reliability,} trust, and human factors---govern the reliability and effectiveness of the cooperative safety loop.

\begin{figure}[t]
    \centering    \includegraphics[width=0.8\linewidth]{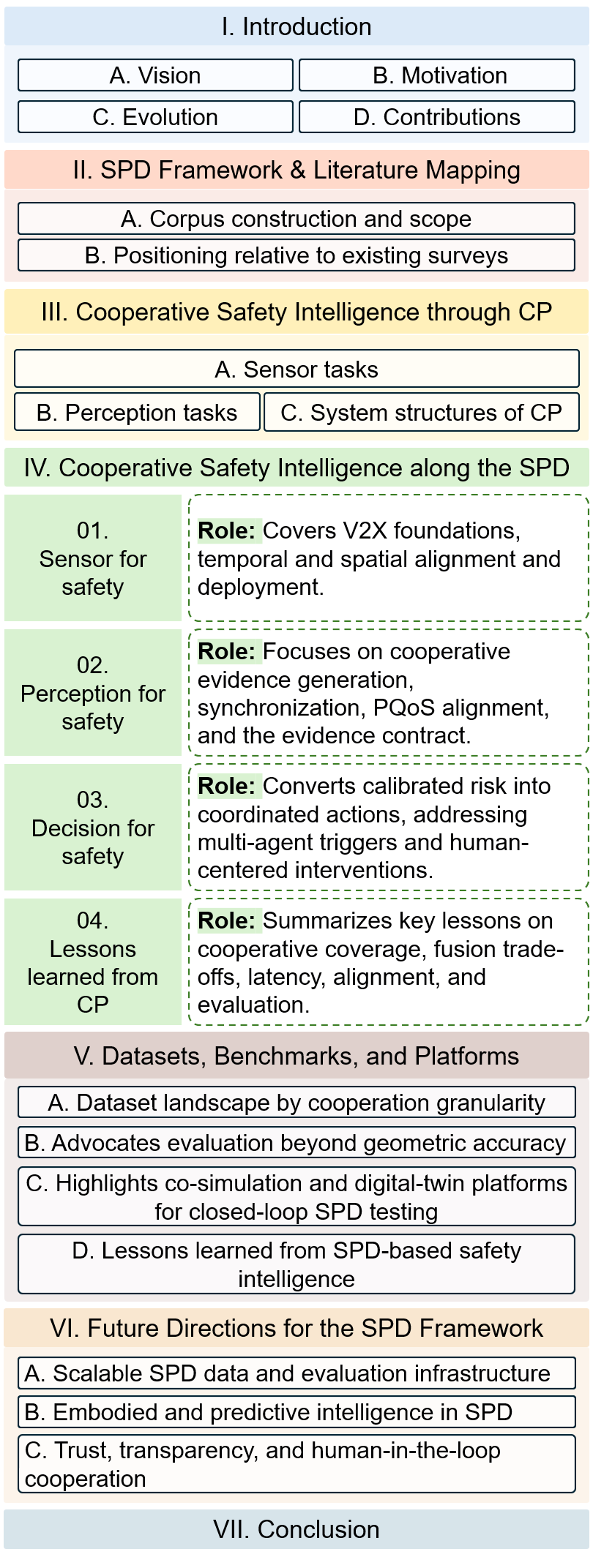}
    \caption{Road map of the survey.}
    \label{fig:roadmap}
    \vspace{-5mm}
\end{figure}

By synthesizing these dimensions, the survey addresses three guiding questions: 
(i) what constitutes safety-ready evidence in cooperative environments, 
(ii) how such evidence can be maintained and calibrated to sustain trustworthy decision-making, and 
(iii) how predictive signals evolve into executable interventions across drivers, vehicles, and infrastructure under different operational conditions.
Building on this framing, our work makes four main contributions:
\begin{itemize}

  \item {We propose a closed-loop SPD organizing framework that maps raw signals into actionable safety intelligence through layer-wise states, interfaces, and feedback principles. It connects sensing, cooperative perception, communication constraints, infrastructure roles, and decision mechanisms, while characterizing safety-ready evidence through risk, timing, Perception Quality of Service (PQoS), and coordination.}

  \item {We construct a structured review corpus of V2X-enabled safety studies from 2017--2025 and organize the included works according to the proposed SPD framework, mapping them into sensing, perception, decision, and cross-layer cooperation categories to support the subsequent technical synthesis.}
  
  \item {We synthesize cooperative perception foundations (sensor tasks, fusion regimes, organizational
  structures, and communication-aware constraints) and relate them along the SPD loop, emphasizing
  timeliness/synchronization, PQoS-aware evidence contracts, and human-centered
  intervention with graded alerts and graceful degradation/recovery.}
  
  \item We organize datasets by cooperation granularity (single-vehicle, V2V, V2I/RSU,
  hybrid), articulate evaluation that goes beyond geometry (earliness, calibration,
  coordination), and discuss co-simulation/digital-twin platforms that record
  PQoS/synchronization for auditable end-to-end safety. We conclude with a roadmap
  for scalable data infrastructures, embodied predictive intelligence, and
  trustworthy human-in-the-loop cooperation.
\end{itemize}

To guide readers through these contributions, Fig.~\ref{fig:roadmap} outlines the overall structure of the survey. 
It shows how the paper progresses from background and foundational concepts through the three layers of the \textit{SPD} framework, followed by the organization of existing studies from partial cooperation to holistic SPD-oriented perspectives, and then toward case studies, open challenges, and concluding insights.

\section{Development of the SPD Unified Framework and Literature Mapping}

This section presents a unified SPD framework for organizing and analyzing V2X-enabled safety intelligence, where safety is understood as the progressive interaction among sensing, perception, and decision layers. {The framework provides a conceptual and methodological structure for mapping broad cooperative research topics---including sensing assets, cooperative fusion, edge infrastructure, communication constraints, trust, human factors, and risk intervention---into SPD states, interfaces, and feedback paths.}

\subsection{Corpus Construction and Scope}

{A structured database search was used to construct the review corpus. The search combined V2X cooperation terms with SPD-aligned technical terms covering sensing assets, cooperative perception, motion forecasting, risk-aware decision-making, synchronization, latency compensation, bandwidth adaptation, and cross-layer coordination.}

{The corpus focuses on vehicle- and traffic-safety studies in which V2X communication or networking mechanisms support safety-oriented functions such as cooperative perception, motion forecasting, risk assessment, warning, and coordinated intervention. Communication-centered studies are included when their mechanisms are explicitly linked to vehicle-safety outcomes. After records were merged, duplicates were removed primarily by DOI and title. The remaining papers were screened by title, abstract, and full text to confirm V2X cooperation and safety relevance. Each included work was mapped to sensing, perception, decision, or cross-layer cooperation according to its primary technical focus.}

\subsection{Positioning Relative to Existing Surveys}

Table~\ref{tab:v2x_survey_related} positions this survey relative to representative studies on V2X-related safety. The comparison is organized according to the dimensions emphasized by the proposed SPD framework, including sensing, perception, decision, cooperation, and safety intelligence.
Earlier reviews primarily focus on specific functional layers or application domains. For instance, Sarker \textit{et~al.}~\cite{sarker2019review} and Fu \textit{et~al.}~\cite{fu2021survey} emphasized sensing and communication infrastructures, while Sun \textit{et~al.}~\cite{sun2023toward} and Karle \textit{et~al.}~\cite{karle2022scenario} concentrated on perception safety, scene understanding, and prediction. Other works, such as Deng \textit{et~al.}~\cite{deng2023survey} and Tan \textit{et~al.}~\cite{tan2024beam}, explored networked control or communication design, yet remained bounded to single-layer or signal-level perspectives. More recent studies---for example, Gao \textit{et~al.}~\cite{gao2024survey}, Abdi \textit{et~al.}~\cite{abdi2024advancing}, and Tan \textit{et~al.}~\cite{tan2025multi}---have begun to bridge sensing and perception, especially through intersection fusion and multi-modal integration, but still lack unified modeling of decision and cooperation mechanisms.

Compared with these works, the present survey establishes a holistic view of {Cooperative Safety Intelligence} under the SPD paradigm. It integrates the three SPD layers---sensing, perception, and decision---with cooperation as a cross-layer organizing dimension, and explicitly connects them to the emergent notion of safety intelligence, where predictive, communicative, and cognitive processes co-evolve toward zero-accident mobility.
{We further provide a concise corpus-construction process that supports transparent screening, scope definition, and SPD-aligned organization of the reviewed V2X-enabled safety literature.}

As reflected by the full-circle marks in Table~\ref{tab:v2x_survey_related}, our synthesis treats cooperation as an organizing principle across the SPD pipeline and extends safety discussion beyond perception reliability to \textit{operational intelligence}. {The combination of reproducible corpus construction and SPD-layer mapping offers a structured basis for future surveys and benchmarks, while consolidating a coherent view of the V2X-enabled safety landscape.}

\begin{table*}[t]
\centering
\caption{Representative Surveys Related to V2X-Enabled Safety Intelligence.}
\label{tab:v2x_survey_related}
\small
\renewcommand{\arraystretch}{1.3}
\setlength{\tabcolsep}{4pt}
\rowcolors{2}{yellow!10}{yellow!3}

\begin{tabular*}{0.99\textwidth}{@{}
c
c
p{7.2cm}
>{\centering\arraybackslash}p{1.0cm}
>{\centering\arraybackslash}p{1.3cm}
>{\centering\arraybackslash}p{1.0cm}
>{\centering\arraybackslash}p{1.6cm}
>{\centering\arraybackslash}p{1.6cm}
@{}}
\hline
\rowcolor{yellow!25}
\textbf{Reference} & \textbf{Year} & \textbf{Survey Focus} &
\textbf{Sensor} & \textbf{Perception} & \textbf{Decision} &
\textbf{Cooperation} & \textbf{Safety Intelligence} \\
\hline

\cite{sarker2019review} & 2019 &
Safety frameworks of CAVs and cooperative ITS &
\halfcircle & \halfcircle & \emptycircle & \halfcircle & \halfcircle \\ 

\cite{tan2021human} & 2021 &
Human--machine interaction in CAVs &
\halfcircle & \fullcircle & \halfcircle & \halfcircle & \halfcircle \\

\cite{fu2021survey} & 2021 &
Driving safety combining sensing, V2X, and AI-based collision avoidance &
\fullcircle & \fullcircle & \halfcircle & \halfcircle & \halfcircle \\

\cite{karle2022scenario} & 2022 &
Scenario understanding and motion prediction models for autonomous vehicles &
\halfcircle & \fullcircle & \halfcircle & \emptycircle & \halfcircle \\

\cite{sun2023toward} & 2023 &
Perception safety and SOTIF standardization in autonomous driving &
\halfcircle & \fullcircle & \emptycircle & \emptycircle & \halfcircle \\

\cite{deng2023survey} & 2023 &
Communication--control co-design for networked motion control safety &
\halfcircle & \halfcircle & \fullcircle & \halfcircle & \halfcircle \\

\cite{gao2024survey} & 2024 &
Collaborative perception at intersections (RSU--vehicle fusion) &
\fullcircle & \fullcircle & \emptycircle & \fullcircle & \halfcircle \\

\cite{tan2024beam} & 2024 &
Beam alignment and mmWave communication for V2X &
\fullcircle & \halfcircle & \emptycircle & \halfcircle & \emptycircle \\

\cite{oghorada2024benefits} & 2024 &
Benefits of V2X communication for cooperative and automated driving &
\halfcircle & \halfcircle & \halfcircle & \fullcircle & \fullcircle \\

\cite{yusuf2024vehicle} & 2024 &
Technical review of V2X, sensors, and AI for VRU safety in autonomous vehicles &
\fullcircle & \fullcircle & \halfcircle & \fullcircle & \fullcircle \\

\cite{abdi2024advancing} & 2024 &
Interdisciplinary review on V2X communication and trajectory prediction for VRU protection &
\fullcircle & \fullcircle & \halfcircle & \halfcircle & \fullcircle \\

\cite{tian2024recent} & 2025 &
Vehicle--road--pedestrian state estimation and fusion methods &
\fullcircle & \fullcircle & \halfcircle & \halfcircle & \fullcircle \\

\cite{tan2025multi} & 2025 &
Multi-modal sensing and ISAC integration for intelligent V2X &
\fullcircle & \fullcircle & \halfcircle & \fullcircle & \fullcircle \\

\rowcolor{yellow!20}
{This Survey} & {2026} &
{Cooperative safety intelligence over V2X networks} &
\textbf{\fullcircle} & \textbf{\fullcircle} & \textbf{\fullcircle} &
\textbf{\fullcircle} & \textbf{\fullcircle} \\
\hline
\end{tabular*}

\vspace{0.5mm}
\begin{tabular}{@{}p{0.99\textwidth}@{}}
\rowcolor{yellow!10}
\footnotesize
\textbf{Note:}
\fullcircle~=~Primary coverage;
\halfcircle~=~Partial coverage;
\emptycircle~=~Limited coverage.
{Our survey provides an integrated view across sensing, perception, decision, and cooperation layers, and organizes V2X-enabled safety studies through structured corpus construction and SPD-aligned mapping.}
\\
\hline
\end{tabular}

\end{table*}

\section{Toward Cooperative Safety Intelligence through Cooperative Perception}

{The advancement of cooperative safety intelligence in V2X-enabled systems can be understood as an evolutionary process from data-level sharing to perception-level understanding and further to decision-level coordination~\cite{gao2024vehicle,wei2025cooperative}. At its foundation, cooperative perception merges distributed observations from vehicles and infrastructure, thereby extending individual sensing into a collective representation of the traffic environment~\cite{caillot2022survey,xiang2023multi}. Building upon this capability, the SPD framework broadens the role of cooperative perception from shared scene understanding to safety-ready evidence generation, shared risk assessment, and coordinated intervention within a closed-loop sensing--perception--decision process~\cite{lv2022cooperative,fu2023cooperative}.}

\subsection{Sensor Tasks in Cooperative Perception}

The foundation of cooperative perception lies in the sensing layer, which integrates diverse modalities to capture complementary aspects of the environment. Each sensing principle contributes a trade-off between precision, range, robustness, and bandwidth efficiency. By linking heterogeneous sensors across vehicles, RSUs, and potentially aerial nodes, cooperative perception establishes a distributed perceptual field that transcends the limitations of single-agent sensing. This section provides an overview of sensor modalities currently employed in cooperative perception, covering their cooperative roles, operational mechanisms, and evolving integration trends.

\subsubsection{Camera-based Sensing}
Vision sensors remain a highly informative and cost-effective modality for environmental understanding. Cameras provide dense appearance, color, and texture information that are critical for object classification, traffic light recognition, and semantic segmentation~\cite{zhong2022empowering}. Multi-camera systems, including stereo, fisheye, and panoramic arrays, enable wide-angle coverage and depth estimation through epipolar geometry. In cooperative perception, vehicle-mounted and RSU-mounted cameras are often paired to overcome occlusions and extend the visual field. The use of BEV transformation and deep multi-view fusion further enables consistent spatial reasoning across agents. Recent literature highlights the trend toward \emph{multi-camera cooperative networks}, where asynchronous visual streams from multiple nodes are temporally aligned and semantically fused to build unified scene graphs~\cite{nabati2021centerfusion,shan2024experimental}. Nevertheless, challenges remain in illumination robustness, weather sensitivity, and long-range accuracy, motivating integration with active sensors for redundancy and cross-validation.

\subsubsection{LiDAR-based Sensing}
LiDAR (Light Detection and Ranging) delivers high-resolution 3D geometry and precise depth information indispensable for object localization and motion forecasting. Cooperative LiDAR setups exploit multi-view point cloud aggregation to compensate for occlusion and sparsity~\cite{wu2020improved}. Multi-layer spinning LiDARs provide dense 360° coverage, while solid-state LiDARs offer lower cost and faster refresh rates. The fusion of LiDAR point clouds from different agents poses alignment challenges due to varying poses, timestamps, and sampling densities~\cite{zhang2020gc}. To address these, voxel-based registration and keypoint compression are employed before V2X transmission~\cite{tan2025multi}. Datasets such as V2X-Sim, OPV2V\cite{xu2022opv2v}, and V2X-Seq demonstrate that sharing LiDAR-based intermediate representations significantly improves 3D detection performance in cooperative settings~\cite{li2022v2x}. The ongoing shift toward \emph{cooperative LiDAR sensing} transforms each agent from a passive observer into an active contributor to shared geometric context.

\subsubsection{Radar and Millimeter-Wave-based Sensing}
Radar sensors measure both range and velocity via Doppler shifts and are particularly effective under adverse weather or poor visibility. Their immunity to fog, rain, and snow makes them a stable complement to optical sensors. Modern 77~GHz and 120~GHz mmWave radars achieve centimeter-level range resolution and enhanced angular precision through frequency-modulated continuous wave (FMCW) and multiple-input multiple-output (MIMO) architectures, while remaining sensitive to mobility-induced beam misalignment in V2X scenarios~\cite{nabati2021centerfusion,shan2024experimental,tan2024beam}. Cooperative radar perception leverages shared radar cross-section (RCS) features and occupancy grids among vehicles to estimate the motion of hidden objects. 
{Recent studies have highlighted Integrated Sensing and Communication (ISAC) as a promising direction for V2X systems, where shared spectrum, waveform, or hardware resources jointly support radar-like sensing and vehicular data exchange~\cite{noor20226g,ko2021v2x}. In 6G V2X, ISAC is particularly relevant to safety-critical scenarios because sensing latency, detection reliability, beam alignment, and communication QoS need to be jointly optimized within the same cooperative loop~\cite{wei2025low,liu2026decentralized}. This integration can improve spectrum utilization and provide timely environmental awareness for cooperative safety applications.}

\subsubsection{Ultrasonic, Infrared, and Thermal-based Sensing}
Although often overlooked in high-level perception, low-cost ultrasonic and infrared sensors play an important supporting role in cooperative perception. Ultrasonic sensors detect nearby obstacles and are particularly effective for low-speed maneuvering or blind-zone monitoring. Infrared and thermal cameras provide valuable cues under poor lighting or night-time conditions, enhancing pedestrian and cyclist detection~\cite{tan2025multi}. Large-scale RSU deployments may integrate such lightweight sensors to establish redundant safety coverage at low cost. Hybrid vision-thermal fusion frameworks have demonstrated improved performance in multi-agent low-visibility environments, emphasizing the practical benefit of cross-spectrum cooperation.

\subsubsection{GNSS and IMU-based Sensing and Positioning}
Global Navigation Satellite System (GNSS) and Inertial Measurement Unit (IMU) jointly provide spatiotemporal referencing for distributed agents, ensuring that observations can be transformed into a common coordinate system. High-precision Real-Time Kinematic (RTK) based-GNSS achieves centimeter-level localization accuracy, while IMU provides short-term pose estimation between GNSS updates~\cite{zhuang2021cooperative,zhang2024road}. Cooperative perception frameworks typically fuse GNSS/IMU information with LiDAR-based SLAM to maintain alignment between moving agents and static RSUs~\cite{li2024nlos}. However, urban canyons and tunnels cause signal blockage or multipath effects, reducing reliability. To mitigate these issues, hybrid approaches integrate Ultra-Wideband (UWB) or Dedicated Short-Range Communications (DSRC) ranging to provide fallback positioning, thus ensuring continuity of cooperative localization even in GNSS-denied environments.

\subsubsection{RF-based Sensing and Positioning}
Radio Frequency (RF)-based methods utilize existing communication signals for localization and environmental awareness. These include Wi-Fi, Cellular-V2X (C-V2X), and emerging 5G/6G systems that support sub-meter ranging precision. The major techniques can be categorized as follows:  
\textit{(a) RSSI (Received Signal Strength Indication)} measures distance through path-loss models but is susceptible to multipath fading.  
\textit{(b) TOA (Time of Arrival)} and \textit{TDOA (Time Difference of Arrival)} estimate distance based on signal propagation time, offering higher accuracy under synchronized conditions.  
\textit{(c) AOA (Angle of Arrival)} leverages antenna arrays to infer incident direction and has been widely used for triangulation in V2X networks.  
\textit{(d) Fingerprinting} matches real-time channel characteristics against a pre-built radio map to achieve data-driven localization.  
{Recent developments in DSRC, LTE-V2X, 5G NR-V2X, and emerging 6G-oriented systems have fostered the convergence of communication and sensing, where RF signals are increasingly used for both connectivity and situational awareness~\cite{moradi2023dsrc,garcia2021tutorial}.}
In this \emph{communication-aware sensing} paradigm, RF-based features are shared among agents to detect Non-Line-of-Sight (NLOS) objects or infer motion trajectories, complementing traditional geometric modalities.

\begin{table*}[t]
\caption{Representative Sensor-Layer Cooperative Perception Datasets.}
\label{tab:coop_datasets}
\centering
\small
\setlength{\tabcolsep}{4pt}
\renewcommand{\arraystretch}{1.2}
\rowcolors{2}{yellow!10}{yellow!3}
\resizebox{1\textwidth}{!}{
\begin{tabular}{
    l l p{2cm} c ccc ccc c c
}
\rowcolor{yellow!25}
\hline
\textbf{Dataset} & \textbf{Publication} & \textbf{Source} & \textbf{Scenario} &
\multicolumn{3}{c}{\textbf{Sensors}} &
\multicolumn{3}{c}{\textbf{Tasks}} &
\textbf{LiDAR Frames} & \textbf{Views} \\
\cline{5-7} \cline{8-10}
\rowcolor{yellow!25}
 & & & & \textbf{RGB} & \textbf{Depth} & \textbf{LiDAR} & \textbf{Det.} & \textbf{Track.} & \textbf{Seg.} & & \\
\hline
\rowcolor{gray!15}
\multicolumn{12}{l}{\textit{Source Type: Simulation-Based Datasets}} \\
VANETs~\cite{maalej2017vanets} & GLOBECOM-2017 & KITTI & V2V & \checkmark &  & \checkmark & \checkmark &  &  & -- & -- \\
V2V-Sim~\cite{wang2020v2vnet} & ECCV-2020 & LiDARsim & V2V &  &  & \checkmark & \checkmark &  & \checkmark & 51,200 & -- \\
CoopInf~\cite{arnold2020cooperative} & TITS-2020 & CARLA & V2I & \checkmark &  &  & \checkmark &  &  & 10,000 & 6,8 \\
V2X-Sim~\cite{li2022v2x} & RAL-2022 & CARLA+SUMO & V2V/V2I & \checkmark & \checkmark & \checkmark & \checkmark &  & \checkmark & 10,000 & 2–5 \\
OPV2V~\cite{xu2022opv2v} & ICRA-2022 & CARLA+OpenCDA & V2V & \checkmark &  & \checkmark & \checkmark & \checkmark & \checkmark & 11,464 & 2–7 \\
AUTOCASTSIM~\cite{cui2022coopernaut} & CVPR-2022 & CARLA & V2V &  &  & \checkmark & \checkmark &  &  & -- & -- \\
V2XSet~\cite{xu2022v2x} & ECCV-2022 & CARLA+OpenCDA & V2V/V2I & \checkmark &  & \checkmark & \checkmark & \checkmark &  & 11,447 & 2–5 \\
CARTI~\cite{pillargrid2022} & ITSC-2022 & CARLA & V2I &  &  & \checkmark & \checkmark & \checkmark &  & 11,000 & 2 \\
DeepAccident~\cite{wang2024deepaccident} & AAAI-2024 & CARLA & V2V/V2I & \checkmark &  & \checkmark & \checkmark & \checkmark & \checkmark & 57,000 & 5 \\
\rowcolor{gray!15}
\multicolumn{12}{l}{\textit{Source Type: Real-World Datasets Collected from On-Road Multi-Agent Scenarios}} \\
T\&J~\cite{chen2019cooper} & ICDCS-2019 & Real-World & V2V &  &  & \checkmark & \checkmark &  &  & 100 & 2 \\
WIBAM~\cite{howe2021weakly} & BMVC-2021 & Real-World & V2I & \checkmark &  &  & \checkmark & \checkmark & \checkmark & 33,092 & 2–4 \\
DAIR-V2X-C~\cite{yu2022dair} & CVPR-2022 & Real-World & V2I & \checkmark &  & \checkmark & \checkmark & \checkmark &  & 38,845 & 2 \\
V2V4Real~\cite{xu2023v2v4real} & CVPR-2023 & Real-World & V2V & \checkmark &  & \checkmark & \checkmark & \checkmark &  & 20,000 & 2 \\
V2X-Seq~\cite{yu2023v2x} & CVPR-2023 & Real-World & V2V/V2I & \checkmark &  & \checkmark & \checkmark & \checkmark &  & 15,000 & 2 \\
Tumtraf V2X~\cite{Zimmer2024Tumtraf} & CVPR-2024 & Real-World & V2I & \checkmark & \checkmark & \checkmark & \checkmark & \checkmark &  & 2,000 & 9 \\
V2X-Real~\cite{xiang2024v2xreal} & ECCV-2024 & Real-World & V2V/V2I & \checkmark & \checkmark & \checkmark & \checkmark &  &  & 33,000 & 12 \\
V2X-Radar~\cite{yang2026v2x} & {NeurIPS-2025} & Real-World & V2V/V2I & \checkmark & \checkmark & \checkmark & \checkmark &  &  & 4D(20,000),3D(20,000) & 4 \\
V2X-PnP~\cite{zhou2025v2xpnp} & ICCV-2025 & Real-World & V2V/V2I/I2I & \checkmark &  & \checkmark & \checkmark & \checkmark &  & 40,000 & 4 \\
V2X-ReaLO~\cite{xiang2025v2x} & {arXiv preprint-2025} & Real-World & V2V/V2I/I2I & \checkmark &\checkmark  & \checkmark &  &  &  & 25,028 & 4 \\
\hline
\end{tabular}%
}
\vspace{0.1pt}
\noindent
\begin{tabular}{p{\dimexpr\textwidth-2\tabcolsep\relax}}
\rowcolor{yellow!10}
\footnotesize
\textbf{Note:}
\checkmark~=~Available or supported;
--~=~Not reported.
\textbf{RGB}/\textbf{Depth}/\textbf{LiDAR} columns denote sensor modalities, while \textbf{Det.} (Detection), \textbf{Track.} (Tracking), and \textbf{Seg.} (Segmentation) indicate supported perception tasks.
Frame and view counts approximate dataset scale and agent diversity.
\\
\hline
\end{tabular}
\end{table*}

\subsubsection{Multi-Modal and Hierarchical Sensing Integration}
The evolution of cooperative perception sensors is moving toward multi-modal, hierarchical, and cooperative integration. At the hardware level, vehicles and RSUs combine active and passive sensors to achieve robustness under all-weather, all-visibility conditions. At the algorithmic level, shared embeddings and latent-space feature fusion enable semantic consistency across modalities. Multi-modal fusion also supports uncertainty calibration, allowing downstream modules to weigh sensor contributions based on reliability scores. From a system perspective, cooperative perception forms a three-tier hierarchy:  
\emph{(i)} the upper layer (e.g., cameras and LiDAR) provides high-level semantic cues,  
\emph{(ii)} the middle layer (e.g., radar and GNSS) maintains geometric and temporal continuity, and  
\emph{(iii)} the bottom layer (e.g., RF and ultrasonic) offers redundancy for safety assurance.  
This multi-scale hierarchy creates a resilient and extensible foundation for the SPD framework, ensuring that what is perceived locally can be contextualized globally and shared cooperatively~\cite{bai2024survey,zha2025heterogeneous}. 

Table \ref{tab:coop_datasets} summarizes how dataset composition and cooperation topology have co-evolved over recent years. For comprehensive dataset-specific descriptions, readers may refer to surveys dedicated to cooperative perception datasets \cite{wang2025colldataset}. Given the rapid pace of development, our discussion emphasizes influential releases from 2022 onward, with particular attention to datasets introduced in 2024–2025.

Several trends emerge. Early datasets (e.g., KITTI-derived VANETs and MFSL variants) primarily serve as single-modality proofs of concept, often relying on synthetic or loosely coupled multi-agent setups. More recent benchmarks (e.g., OPV2V, V2X-Seq, V2V4Real) incorporate real RSU–vehicle configurations, extended temporal sequences, and richer cross-view geometry, enabling systematic study of occlusion mitigation, temporal synchronization, and multi-view consistency. This evolution supports progressively stronger sensor-layer evaluation and motivates the need for modality-aware cooperative benchmarks that operate under realistic timing constraints.

Overall, the sensor modalities in cooperative perception collectively define the observable scope of cooperative intelligence. Their complementarity—spanning optical, geometric, inertial, and electromagnetic domains—transforms individual sensing into a distributed, multi-perspective process. As sensing technologies converge with communication and computation, the cooperative perception ecosystem is evolving from sensor fusion to \emph{perceptual federation}, wherein each node contributes dynamically to a shared, synchronized, and safety-ready environmental view.

\subsection{Perception Tasks in Cooperative Perception}

\begin{figure*}[t]
    \centering
    \includegraphics[width=0.96\textwidth]{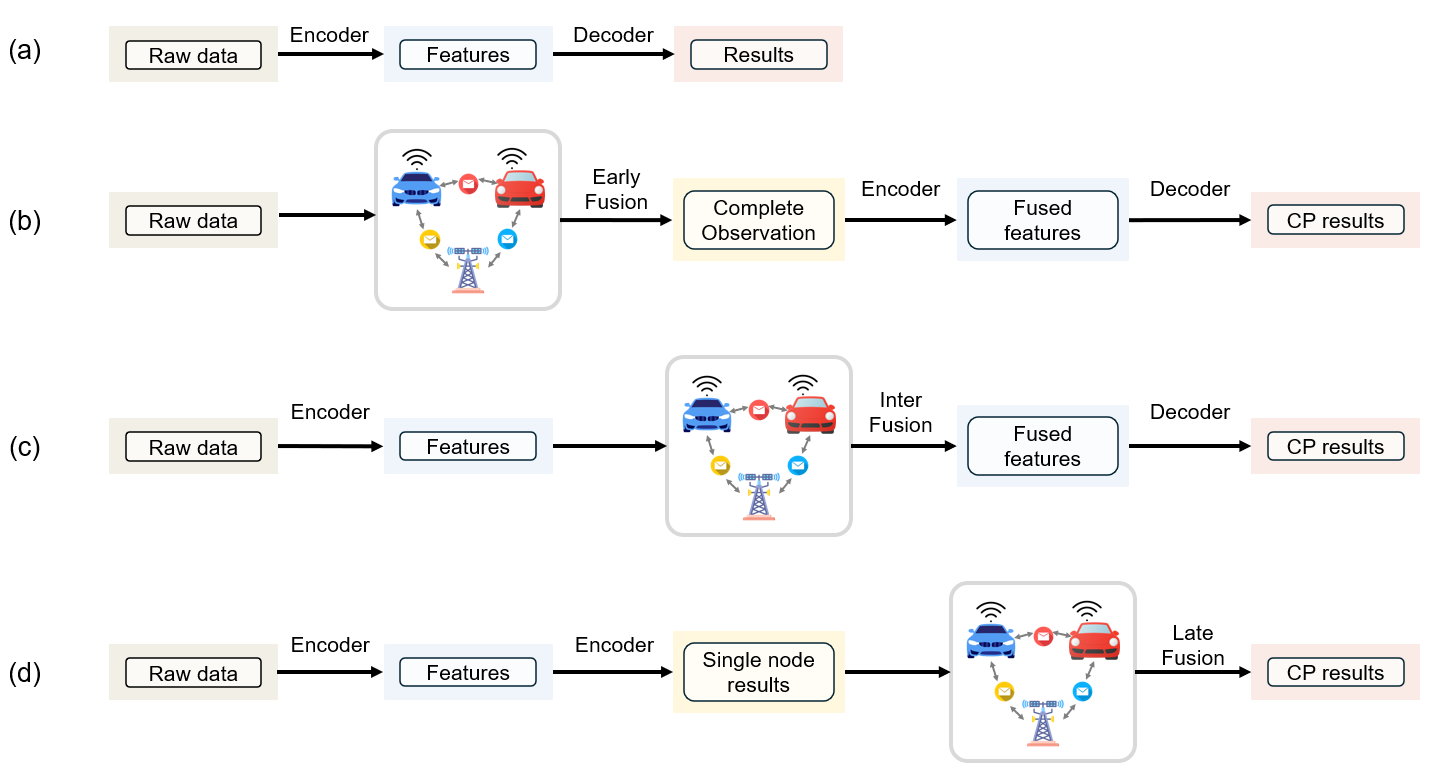}
    \vspace{-6pt}
    \caption{\textbf{Collaboration schemes for cooperative perception.}
    (a) \emph{Single-vehicle baseline:} raw data are encoded into features and decoded into results without collaboration.
    (b) \emph{Early fusion:} agents first exchange raw observations to form a more complete scene, then a shared encoder–decoder produces cooperative perception results.
    (c) \emph{Intermediate fusion:} each agent encodes its sensor data locally, exchanges latent features, and decodes fused features into cooperative perception results.
    (d) \emph{Late fusion:} each agent produces local perception results, which are then exchanged and merged to obtain cooperative perception results.
    The gray panel depicts the communication topology (V2V/V2I), and arrows indicate the processing flow from sensing to fusion and outputs.}
    \label{fig:cp_fusion}
    \vspace{-5mm}
\end{figure*}

\begin{table*}[t]
\caption{Overview of Multisensor Fusion Methods and Their Applications.}
\label{tab:cp_fusion_methods}
\centering
\small
\setlength{\tabcolsep}{4pt}         
\renewcommand{\arraystretch}{1.2}  

\newcolumntype{L}[1]{>{\raggedright\arraybackslash}p{#1}}
\newcolumntype{C}[1]{>{\centering\arraybackslash}p{#1}}

\rowcolors{3}{yellow!10}{yellow!3}

\resizebox{\textwidth}{!}{%
\begin{tabular}{
    l  
    c  
    c  
    C{0.8cm}  
    C{0.8cm}  
    C{0.8cm}  
    C{0.8cm}  
    C{0.8cm}  
    L{1.5cm}  
    C{2cm}  
    L{2.8cm}  
}
\hline
\rowcolor{yellow!25}
\multicolumn{3}{c}{} &
\multicolumn{5}{c}{\textbf{Collaboration Level}} &
\multicolumn{3}{c}{} \\
\rowcolor{yellow!25}
\textbf{Method} & \textbf{Year} & \textbf{Scenario} &
\textbf{Early} & \multicolumn{3}{c}{\textbf{Intermediate}} & \textbf{Late} &
\textbf{Sensors} & \textbf{Dataset} & \textbf{Task} \\
\cline{5-7}
\rowcolor{yellow!25}
 &  &  &  & \textbf{Trad} & \textbf{Attn} & \textbf{Topo} &  &  &  &  \\
\hline

Cooper~\cite{chen2019cooper} &  ICDCS-2019 & V2V & \checkmark &  &  &  &  & LiDAR & KITTI, T\&J  & 3D Object Detection \\
F-Cooper~\cite{chen2019f} & SEC-2019 & V2V &  & \checkmark &  &  &  & LiDAR & KITTI, T\&J & 3D Object Detection \\
PillarGrid~\cite{pillargrid2022} & ITSC-2022 & V2I &  & \checkmark &  &  &  & LiDAR & CARTI & 3D Object Detection \\
COOPERNAT~\cite{cui2022coopernaut} & CVPR-2022 & V2V &  &  & \checkmark &  &  & LiDAR &  AUTOCASTSIM framework & End-to-End Driving \\
V2X-ViT~\cite{xu2022v2x} & ECCV-2022 & V2X &  &  & \checkmark & \checkmark &  & LiDAR & V2XSet & 3D Object Detection \\
CoCa3D~\cite{coca3d2023} & CVPR-2023 & V2I,V2U &  & \checkmark &  &  &  & UAV Camera &  OpenV2V, DAIR-V2X, CoPerception-UAVs & 3D Object Detection \\
V2X-ViT2~\cite{xu2024v2x} & TPAMI-2024 & V2X &  &  & \checkmark &  \checkmark &  & LiDAR & V2XSet & 3D Object Detection \\
UniV2X~\cite{univ2x2024} & AAAI-2024 & V2I &  & \checkmark &  &  &  & Camera & DAIR-V2X & Trajectory Planning \\
MRCNet~\cite{hong2024multi} & CVPR-2024 & V2X &  &  & \checkmark & \checkmark & \checkmark & LiDAR & DAIR-V2X & 3D Object Detection (noise-robust) \\
HEAL~\cite{lu2024heal} & ICLR-2024 & V2X &  &  & \checkmark &  & & Lidar, Camera & OPV2V-H & 3D Object Detection (heterogeneous) \\

HDAAGT~\cite{Abdi2025HDAAGT} & TITS-2025 & V2I &  &  & \checkmark &  & & Fisheye Camera & Fisheye-MARC & Trajectory prediction \\
\hline
\end{tabular}%
}
\vspace{0.1pt}
\noindent
\begin{tabular}{p{\dimexpr\textwidth-2\tabcolsep\relax}}
\rowcolor{yellow!10}
\footnotesize
\textbf{Note:} 
\checkmark~=~Available or supported; 
\textbf{Trad}: Traditional Fusion; 
\textbf{Attn}: Attention-Based Fusion; 
\textbf{Topo}: Topology-Based Fusion.
\\
\hline
\end{tabular}
\end{table*}

The perception layer of cooperative perception transforms distributed sensory inputs into structured, semantically meaningful representations of the environment. 
It extends beyond raw sensing to include multi-agent feature extraction, alignment, and fusion, thereby enabling vehicles and infrastructure to achieve a collective understanding of dynamic scenes. Through cooperation, agents can detect occluded participants, infer interactions, and construct shared world models that are both spatially and temporally consistent. Depending on the stage at which information is exchanged, cooperative perception systems can be broadly categorized into \textit{early}, \textit{intermediate}, and \textit{late} fusion paradigms~\cite{wei2025cooperative}. 
Each paradigm reflects a distinct trade-off between communication cost, fusion accuracy, and temporal synchronization, as summarized in Table~\ref{tab:cp_fusion_methods} of representative approaches.

As shown in Fig.~\ref{fig:cp_fusion}, cooperative perception pipelines differ mainly in \emph{when} and \emph{what} to share across agents. Early fusion aggregates raw sensory data before encoding, yielding complete but communication-heavy observations. Intermediate fusion exchanges latent features, balancing bandwidth and robustness while enabling cross-view reasoning. Late fusion transmits final detection results, requiring minimal bandwidth but losing fine-grained alignment.

{Beyond the fusion stage itself, cooperative perception methods also differ in how they handle system-level V2X constraints, including bandwidth consumption, communication latency, packet loss, NLOS coverage, pose misalignment, and heterogeneous deployment. Table~\ref{tab:comm_efficiency_timeliness} summarizes representative strategies under these constraints and highlights coupled evaluation across perception quality, communication efficiency, robustness, and deployment feasibility. Compression, selective communication, and agent selection improve communication efficiency by reducing redundant data exchange, while latency-compensation and lossy-communication methods enhance robustness under asynchronous or interrupted links. NLOS/blind-zone compensation, pose correction, and heterogeneous fusion further expand cooperative coverage and interoperability, with additional requirements on calibration, temporal modeling, alignment, or infrastructure support.}

{While Table~\ref{tab:comm_efficiency_timeliness} summarizes system-level method trade-offs, recent cooperative perception studies are evaluated under heterogeneous benchmark protocols. Datasets, sensor configurations, cooperation modes, agent numbers, IoU thresholds, class settings, communication budgets, delay models, and task definitions often differ across works. Therefore, directly aggregating numerical values such as AP/mAP, latency, bandwidth usage, communication overhead, compression ratio, and NLOS robustness into a single leaderboard may obscure protocol differences and imply unsupported head-to-head comparability. To provide concrete evaluation evidence while preserving fair interpretation, Table~\ref{tab:benchmark_evidence_recent_cp} summarizes representative benchmark studies together with their reported metrics or settings, benchmark evidence, and relevance to SPD-based cooperative safety intelligence.}

{Together, Tables~\ref{tab:comm_efficiency_timeliness} and~\ref{tab:benchmark_evidence_recent_cp} provide complementary comparative views. The former synthesizes method-level system trade-offs, while the latter links these trade-offs to reported benchmark metrics, evaluation settings, datasets, cooperation modes, and use cases. This organization avoids misleading numerical ranking across incompatible protocols while still preserving concrete evidence on perception accuracy, communication efficiency, bandwidth usage, transmission cost, latency/asynchrony robustness, NLOS or occlusion-related coverage, and deployment-oriented evaluation. It therefore helps assess whether shared perception outputs remain accurate, timely, communication-efficient, robust under degraded conditions, and useful for downstream SPD-based safety reasoning.}

\begin{table*}[t]
\centering
\caption{System-level trade-off comparison of cooperative perception methods under communication efficiency, timeliness, and robustness constraints.}
\label{tab:comm_efficiency_timeliness}
\scriptsize
\setlength{\tabcolsep}{3.5pt}
\renewcommand{\arraystretch}{1.15}

\resizebox{\textwidth}{!}{
\begin{tabular}{
>{\centering\arraybackslash}m{4.3cm}
>{\centering\arraybackslash}m{1.1cm}
>{\centering\arraybackslash}m{4.2cm}
>{\centering\arraybackslash}m{4.0cm}
>{\centering\arraybackslash}m{4.2cm}
}
\hline
\rowcolor{yellow!25}
\textbf{Methods} &
\textbf{Sensor} &
\textbf{Strategy} &
\textbf{Advantage} &
\textbf{Trade-off} \\
\hline

\rowcolor{gray!18}
\multicolumn{5}{l}{\textit{Bandwidth Efficiency / Compression}} \\
\rowcolor{yellow!10}
V2VNet~\cite{wang2020v2vnet}, DiscoGraph~\cite{li2021learning}, AttFusion~\cite{xu2022opv2v}, V2X-ViT~\cite{xu2022v2x}, FFNet~\cite{yu2023vehicle}, VIMI~\cite{wang2023vimi}, ScalCompress~\cite{yuan2023scalable} &
L / C &
Feature compression; channel reduction; scalable bitstream &
Bandwidth $\downarrow$; transmission burden $\downarrow$; accuracy--cost balance $\uparrow$ &
Information loss; weak-cue suppression; degradation under rare scenes \\
\hline

\rowcolor{gray!18}
\multicolumn{5}{l}{\textit{Agent Selection}} \\
\rowcolor{yellow!3}
Who2com~\cite{liu2020who2com}, When2com~\cite{liu2020when2com}, MASH~\cite{glaser2021overcoming} &
C &
Handshake communication; learned communication graph &
Redundant links $\downarrow$; agent-level bandwidth use $\downarrow$ &
Useful-agent omission; unstable selection under occlusion \\
\hline

\rowcolor{gray!18}
\multicolumn{5}{l}{\textit{Selective Communication}} \\
\rowcolor{yellow!10}
Where2comm~\cite{hu2022where2comm}, CoCa3D~\cite{coca3d2023}, Learn2comm~\cite{wang2023collaborative}, How2comm~\cite{yang2023how2comm}, CenterCoop~\cite{zhou2023centercoop}, BM2CP~\cite{zhao2023bm2cp}, UMC~\cite{wang2023umc}, What2comm~\cite{yang2023what2comm} &
L / C &
Region/channel selection; confidence maps; object-centered sharing &
Communication efficiency $\uparrow$; task-relevant evidence preserved &
Relevance-estimation error; uncertainty calibration dependency \\
\hline

\rowcolor{gray!18}
\multicolumn{5}{l}{\textit{Latency Compensation}} \\
\rowcolor{yellow!3}
V2X-ViT~\cite{xu2022v2x}, SyncNet~\cite{lei2022latency}, FFNet~\cite{yu2023vehicle}, CoBEVFlow~\cite{wei2023asynchrony}, How2comm~\cite{yang2023how2comm} &
L / C &
Delay-aware encoding; temporal synchronization; BEV-flow prediction &
Asynchrony robustness $\uparrow$; perception stability $\uparrow$ &
Temporal drift; prediction error; failure under abrupt motion \\
\hline

\rowcolor{gray!18}
\multicolumn{5}{l}{\textit{Lossy Communication}} \\
\rowcolor{yellow!10}
V2VAM+LCRN~\cite{li2023learning}, V2X-INCOP~\cite{ren2024interruption} &
L &
Feature repair; historical recovery; interruption-aware inference &
Robustness to packet loss $\uparrow$; incomplete-feature recovery &
History dependency; degradation under long interruptions \\
\hline

\rowcolor{gray!18}
\multicolumn{5}{l}{\textit{NLOS / Blind-Zone Compensation}} \\
\rowcolor{yellow!3}
APMM+MoHeD~\cite{li2024nlos} &
L &
Blind-zone modeling; relay-node selection; visibility estimation &
BLOS coverage $\uparrow$; occlusion-aware safety evidence $\uparrow$ &
Visibility-model dependency; relay availability; infrastructure placement \\
\hline

\rowcolor{gray!18}
\multicolumn{5}{l}{\textit{Pose Error / Heterogeneous Deployment}} \\
\rowcolor{yellow!10}
RobustV2VNet~\cite{vadivelu2021learning}, FPV-RCNN~\cite{yuan2022keypoints}, CoAlign~\cite{lu2023robust}, FeaCo~\cite{gu2023feaco}, MPDA~\cite{xu2023bridging}, HPL-ViT~\cite{liu2024hpl}, HM-ViT~\cite{xiang2023hm}, DI-V2X~\cite{li2024di} &
L / C &
Pose correction; feature alignment; cross-modal/domain adaptation &
Localization robustness $\uparrow$; interoperability $\uparrow$ &
Alignment overhead; sensor mismatch; domain-shift sensitivity \\

\hline
\end{tabular}
}

\vspace{0.1pt}
\noindent
\begin{tabular}{p{\dimexpr\textwidth-2\tabcolsep\relax}}
\rowcolor{yellow!10}
\footnotesize
\scriptsize
\textbf{Note:} L: LiDAR; C: Camera; $\uparrow$: improvement; $\downarrow$: reduction.
\\
\hline
\end{tabular}
\end{table*}

\begin{table*}[t]
\centering
\caption{Representative benchmark and evaluation dimensions in recent cooperative perception studies.}
\label{tab:benchmark_evidence_recent_cp}
\scriptsize
\setlength{\tabcolsep}{1.6pt}
\renewcommand{\arraystretch}{1.18}

\resizebox{0.99\textwidth}{!}{%
\begin{tabular}{
>{\centering\arraybackslash}m{2.4cm}
>{\centering\arraybackslash}m{1.0cm}
>{\centering\arraybackslash}m{2.25cm}
>{\centering\arraybackslash}m{3.2cm}
>{\centering\arraybackslash}m{2.4cm}
>{\centering\arraybackslash}m{2.15cm}
>{\centering\arraybackslash}m{2.55cm}
>{\centering\arraybackslash}m{3.1cm}
>{\centering\arraybackslash}m{3.25cm}
}
\hline
\rowcolor{yellow!25}
\textbf{Work} &
\textbf{Venue} &
\makecell{\textbf{Scenario /}\\\textbf{Dataset}} &
\makecell{\textbf{Perception}\\\textbf{Accuracy}} &
\makecell{\textbf{Comm. /}\\\textbf{Bandwidth}} &
\makecell{\textbf{Delay /}\\\textbf{Robustness}} &
\makecell{\textbf{Main Evaluation}\\\textbf{Focus}} &
\makecell{\textbf{Reported Comparative}\\\textbf{Evidence}} &
\makecell{\textbf{Relevance to SPD-Based}\\\textbf{Cooperative Safety}\\\textbf{Intelligence}} \\
\hline

\rowcolor{yellow!10}
Where2comm~\cite{hu2022where2comm} &
NeurIPS 2022 &
OPV2V, V2X-Sim, DAIR-V2X, CoPerception-UAVs &
\makecell{AP@0.50 gains;\\+25.81\% OPV2V;\\+1.90\% V2X-Sim;\\+7.70\% DAIR-V2X} &
\makecell{$>100K\times$ OPV2V;\\$55\times$ V2X-Sim;\\$105\times$ DAIR-V2X;\\$5128\times$ UAVs}&
Bandwidth-varied settings &
Bandwidth-efficient cooperative perception &
Perception--communication trade-off under bandwidth constraints; confidence-map-based selective sharing &
Reduces communication load via evidence selection \\
\hline

\rowcolor{yellow!3}
V2X-ViT~\cite{xu2022v2x} &
ECCV 2022 &
V2XSet / V2X cooperative perception &
\makecell{AP@0.5/0.7;\\0.882/0.712 perfect;\\0.836/0.614 noisy} &
-- &
\makecell{100 ms delay;\\pose noise;\\57 ms inference} &
Transformer-based BEV cooperation &
Asynchrony, pose-noise, and heterogeneous V2X robustness tests &
Supports timeliness and pose-alignment analysis \\
\hline

\rowcolor{yellow!10}
OPV2V~\cite{xu2022opv2v} &
ICRA 2022 &
Simulated V2V cooperative perception &
\makecell{AP@0.5/0.7;\\PointPillar: 0.908/0.815;\\VoxelNet: 0.906/0.864} &
\makecell{4096$\times$ compression;\\$\sim$3\% AP drop;\\27 Mbps} &
$\sim$5 ms delay &
Fusion-level benchmarking &
Early/intermediate/late fusion comparison under compression &
Shows fusion-stage and bandwidth trade-offs \\
\hline

\rowcolor{yellow!3}
V2X-Sim~\cite{li2022v2x} &
IEEE RA-L 2022 &
Simulated V2V/V2I cooperative perception &
\makecell{AP@0.5/0.7;\\RSU upper: 77.08/72.57;\\DiscoNet+RSU: 72.87/66.40} &
-- &
Multi-agent settings &
Multi-agent, multi-modal, multi-task benchmark &
Detection, tracking, and segmentation benchmark &
Supports multi-task cooperative sensing evaluation \\
\hline

\rowcolor{yellow!10}
DAIR-V2X~\cite{yu2022dair} &
CVPR 2022 &
Real-world V2I cooperative perception &
\makecell{AP$_{3D}$/AP$_{BEV}$@0.5;\\Late: 41.90/47.96;\\Early: 50.03/53.73} &
\makecell{AB:\\336.16 (late)\\$1.382\times10^6$ (early)}&
Temporal asynchrony &
Vehicle--infrastructure 3D detection &
VIC3D benchmark with asynchrony and transmission-cost settings &
Links infrastructure sensing, timing, and cost \\
\hline

\rowcolor{yellow!3}
V2V4Real~\cite{xu2023v2v4real} &
CVPR 2023 &
Real-world V2V cooperative perception &
\makecell{AP@0.5/0.7;\\Sync: 66.5/36.0;\\Async: 58.6/29.7} &
AM: 0.20 MB &
Sync / Async &
Real-world cooperative detection and tracking &
Real-world detection, tracking, and Sim2Real evaluation &
Evidence for V2V occlusion mitigation and deployment \\
\hline

\rowcolor{yellow!10}
V2X-Seq~\cite{yu2023v2x} &
CVPR 2023 &
Sequential real-world V2I cooperative perception &
-- &
\makecell{BPS:\\$6.2\times10^5$ (0ms)\\$1.2\times10^6$ (200ms)} &
0 ms / 200 ms &
Sequential perception and forecasting &
Tracking and forecasting benchmarks under delay settings &
Extends evaluation to forecasting-oriented SPD evidence \\
\hline

\rowcolor{yellow!3}
CoBEVFlow~\cite{wei2023asynchrony} &
NeurIPS 2023 &
IRV2V, DAIR-V2X, V2XSet &
\makecell{IRV2V AP@0.5/0.7;\\300ms: 0.831/0.757;\\500ms: 0.815/0.687;\\V2XSet AP@0.7:\\0.776 / 0.713} &
-- &
\makecell{300 ms / 500 ms;\\delay, interruption,\\clock mismatch} &
Asynchrony-robust BEV fusion &
Tests delays, interruptions, and clock mismatch &
Supports asynchrony-robust safety perception \\
\hline

\rowcolor{yellow!10}
V2X-Real~\cite{xiang2024v2xreal} &
ECCV 2024 &
Real-world V2X cooperative perception &
\makecell{mAP@0.3/0.5;\\VC: 44.8/35.8;\\IC: 50.3/38.5;\\V2V: 46.5/37.3;\\I2I: 62.7/50.3} &
-- &
\makecell{VC, IC, V2V,\\I2I settings} &
Multi-agent and multi-class V2X benchmark &
VC, IC, V2V, and I2I cooperative perception settings &
Evidence for heterogeneous V2X use cases \\
\hline

\rowcolor{yellow!3}
CoSDH~\cite{xu2025cosdh} &
CVPR 2025 &
OPV2V, V2X-Sim, DAIR-V2X &
\makecell{AP@0.5/0.7;\\OPV2V: 96.75/92.92;\\V2X-Sim: 89.23/86.31;\\DAIR-V2X: 76.47/63.76} &
\makecell{BD$\leq$6.75 Mbps;\\OPV2V: 2.0 Mbps\\V2X-Sim: 1.1 Mbps\\DAIR: 1.4 Mbps} &
Real bandwidth limits &
Accuracy--bandwidth trade-off &
Bandwidth-limited communication-efficient comparison &
Links bandwidth demand with perception accuracy \\
\hline

\rowcolor{yellow!10}
\multicolumn{9}{>{\columncolor{yellow!10}\raggedright\arraybackslash}p{\dimexpr22.3cm+16\tabcolsep\relax}}{%
\textbf{Note:}
I2I: infrastructure-to-infrastructure; VIC: vehicle--infrastructure cooperative; 
VC: vehicle-centric; IC: infrastructure-centric; 
AP: Average Precision; mAP: mean Average Precision; 
BPS: bits per second; AB: Average Byte; AM: Average MegaByte; 
BD: bandwidth demand.
} \\
\hline

\end{tabular}%
}
\vspace{-2mm}
\end{table*}

\subsubsection{Early Fusion: Raw Data Alignment and Multi-View Reconstruction}

Early fusion methods transmit raw sensor data--such as point clouds or camera images---between cooperative agents for joint processing. 
This strategy retains maximal information fidelity and supports fine-grained spatial reasoning, but requires precise calibration and time synchronization across nodes. 
Classical systems such as Cooper~\cite{chen2019cooper} and F-Cooper~\cite{chen2019f} aggregate LiDAR point clouds from multiple vehicles to overcome occlusions and expand the sensing range. 
Subsequent works, including UMC~\cite{wang2023umc} and CooperNAUT~\cite{cui2022coopernaut}, improve efficiency through voxelized compression or semantic clustering, reducing redundant data transfer. 
Despite their advantages in completeness and geometric accuracy, early fusion systems are sensitive to communication delay, pose misalignment, and bandwidth limitations. 
They are thus well suited for tightly coupled V2V settings or small-scale RSU-assisted deployments, where global synchronization can be maintained through high-bandwidth channels. 
These methods lay the foundation for feature-level cooperation by demonstrating that synchronized raw observations can substantially enhance collective visibility.

\subsubsection{Intermediate Fusion: Feature-Level Cooperation and Semantic Sharing}

Intermediate fusion constitutes the mainstream paradigm of cooperative perception. 
In this setting, each agent independently performs local feature extraction before transmitting intermediate representations for fusion. 
This approach balances bandwidth efficiency with semantic richness, as shared features encapsulate high-level cues while remaining compact and transferable. 
Three major research directions can be identified according to the fusion mechanism: traditional concatenation-based fusion, attention-driven fusion, and topology-aware fusion.

\paragraph{Traditional Fusion}
Classical methods such as CoCa3D~\cite{coca3d2023} and CoBEVFusion~\cite{cobevfusion2023} combine locally extracted features via concatenation, convolution, or weighted averaging. 
These techniques achieve notable improvements in 3D object detection and BEV semantic segmentation while maintaining stable inference under varying cooperation topologies. 
However, they rely on spatial alignment priors and often suffer from feature redundancy, particularly when agents observe overlapping regions.

\paragraph{Attention-Based Fusion}
To address redundancy and dynamic relevance, attention-based methods explicitly model inter-agent dependencies. 
V2VFormer~\cite{v2vformer2023}, MKD-Cooper~\cite{mkdcooper2023}, and V2XFormer++~\cite{v2vformerpp2023} employ Transformer architectures to compute cross-agent attention, enabling selective feature aggregation based on contextual importance. 
Such models dynamically emphasize informative agents and suppress noisy data, improving robustness under communication dropout. This paradigm marks a key evolution toward \emph{semantic selectivity}, in which information sharing is guided by learned relevance cues and contextual dependencies.

\paragraph{Topology-Based Fusion}
Topology-aware methods integrate spatial relations or graph structures into the fusion process, treating cooperative agents as nodes within an interaction graph. 
Representative works include V2VNet~\cite{wang2020v2vnet}, HEAT~\cite{mo2022multi}, and V2X-ViT~\cite{xu2022v2x}, which encode adjacency or positional embeddings to propagate information along physically or semantically relevant edges. 
These graph-based designs effectively balance scalability and expressiveness, capturing non-local correlations across heterogeneous agents. 
By modeling the underlying communication topology, such frameworks enhance interpretability and reduce dependency on centralized synchronization.

Intermediate fusion thus represents the core of modern cooperative perception research. 
It achieves a favorable trade-off between accuracy, robustness, and scalability, serving as the de facto standard in both academic benchmarks (e.g., OPV2V, V2X-Sim, V2X-Seq) and real-world systems (e.g., DAIR-V2X, DeepAccident).

\subsubsection{Late Fusion: Decision-Level Aggregation and Consensus Reasoning}

Late fusion approaches share processed object lists or detection results after local perception has been completed.
Each agent performs local inference and subsequently exchanges bounding boxes, trajectories, or semantic attributes, which are then merged through consensus or probabilistic reasoning.
Representative frameworks such as DiscoNet~\cite{li2021learning} employ object association and confidence reweighting to reconcile overlapping predictions from multiple viewpoints.
Compared with early and intermediate fusion, this paradigm imposes low communication cost and provides resilience under network degradation or asynchrony.
Its compact output representation retains less fine-grained scene detail, making the recovery of objects missed by individual agents more dependent on the quality and diversity of local detections.
Late fusion is therefore well suited for resource-constrained or heterogeneous systems, where perception results can be efficiently consolidated to support higher-level decision making.

In summary, perception tasks in cooperative perception evolve along a continuum from raw data sharing to feature-level reasoning and ultimately to decision-level consensus. 
Each paradigm contributes a distinct balance between fidelity, scalability, and robustness. 
Early fusion demonstrates the potential of multi-view reconstruction and benefits from accurate synchronization and sufficient bandwidth support.
Intermediate fusion dominates current research by enabling semantic alignment through attention and topology-aware mechanisms, whereas late fusion provides a lightweight yet robust alternative for distributed inference. 
Together, these mechanisms constitute the cognitive foundation of cooperative perception, allowing connected agents to form a shared, semantically grounded understanding of complex traffic scenes. 
Current cooperative perception frameworks provide a strong foundation for collective observation, and their progression toward coordinated action motivates the broader integration of cooperative perception within the decision-aware SPD framework discussed in the following section.

\subsubsection{Communication-Efficient and Semantic Cooperative Perception}

{In communication-constrained V2X settings, recent cooperative perception studies increasingly focus on selecting task-relevant information for transmission under limited bandwidth and time-varying link conditions. A representative direction is bandwidth-efficient cooperative perception. Where2comm introduces spatial confidence maps to identify perceptually critical regions and selectively transmit sparse but informative features, thereby reducing communication cost while maintaining cooperative detection performance~\cite{hu2022where2comm}. Transformer-based BEV cooperation further advances this direction by modeling inter-agent interactions in a unified spatial representation. For example, V2X-ViT employs heterogeneous multi-agent self-attention and multi-scale window attention to fuse information across vehicles and infrastructure under practical challenges such as asynchrony, pose errors, and agent heterogeneity~\cite{xu2022v2x}, while CoBEVT extends cooperative perception to multi-agent multi-camera BEV semantic representation~\cite{xu2022cooperative}.} {Cooperative safety evaluation therefore benefits from jointly considering perception accuracy, communication load, latency, synchronization error, and message relevance, since these factors determine whether shared evidence remains usable for downstream decisions~\cite{huang2025vehicle,zha2025heterogeneous}. Semantic and task-oriented V2X communications address this requirement by emphasizing context-dependent information that is relevant to the receiver's perception or planning task~\cite{wei2025cooperative,lusvarghi2026search}. From the SPD perspective, semantic communication can be interpreted as relevance-aware evidence selection: the sensor and perception layers identify safety-relevant observations, while the decision layer provides the task context in which such evidence becomes actionable~\cite{ren2024interruption,sheng2025semantic}. Together, semantic V2X communication, PQoS-aware evidence exchange, and bandwidth-efficient cooperation provide a communication-aware path toward scalable cooperative safety intelligence.}

{Another emerging line is asynchronous and latency-robust fusion, which addresses delay-induced temporal misalignment in cooperative messages to prevent stale features from distorting safety-critical risk estimation~\cite{wang2025latency,li2024nlos}. In parallel, ISAC is becoming a key 6G-V2X enabler by integrating radar-like sensing with vehicular data exchange~\cite{wei2025low,liu2026decentralized}. For cooperative safety intelligence, ISAC can provide low-latency environmental updates and communication-aware sensing capabilities, while also introducing design trade-offs among sensing accuracy, detection latency, communication QoS, and resource allocation~\cite{long2024deep,li2024frame}.}

\subsection{Structural Organization of Cooperative Perception Systems}
\label{subsec:cp_structural_organization}

\begin{table*}[t]
\centering
\caption{Comparative summary of representative Cooperative Perception Architectures.}
\label{tab:cp_architectures_extended}
\small
\setlength{\tabcolsep}{3.5pt}
\renewcommand{\arraystretch}{1.2}
\rowcolors{2}{yellow!10}{yellow!3}
\resizebox{\textwidth}{!}{%
\begin{tabular}{
    l  
    c  
    c  
    c  
    c  
    c  
    c  
    c  
    c  
    c  
    c  
}
\rowcolor{yellow!25}
\hline
\textbf{Work} & \textbf{Year} & \textbf{Data Granularity} & \textbf{Modality} &
\textbf{Central} & \textbf{Edge} & \textbf{Fusion} &
\textbf{Scalability} & \textbf{Adaptivity} & \textbf{Efficiency} & \textbf{Evaluation Scope} \\
\hline

\rowcolor{gray!15}
\multicolumn{11}{l}{\textit{Centralized Architectures}} \\
EdgeCooper\cite{luo2023edgecooper} & 2023 & Feat & L & \checkmark & \checkmark & Interm. & \checkmark & \checkmark & \checkmark & Sim+Real \\
MARL\cite{qu2024model} & 2024 & Feat & L & \checkmark & \checkmark & Interm. & \checkmark & \checkmark & \checkmark & Sim \\
Rcooper\cite{hao2024rcooper} & 2024 & Obj & C+L & \checkmark & \checkmark & Late & \checkmark &   & \checkmark & Real \\
RLDOF\cite{linl2025roadside} & 2025 & Feat & L & \checkmark & \checkmark & Early & \checkmark & \checkmark & \checkmark & Real \\
ECOP\cite{hou2025enhancing} & 2025 & Feat & L & \checkmark & \checkmark & Interm. & \checkmark & \checkmark & \checkmark & Sim \\
\rowcolor{gray!15}
\multicolumn{11}{l}{\textit{Decentralized Architectures}} \\
DCP\cite{yoon2021performance} & 2021 & Obj & L &   &   & Late & \checkmark & \checkmark & \checkmark & Sim+Real \\
AIR-DRA\cite{bischoff2021prioritizing} & 2021 & Feat & C+L &   &   & Late & \checkmark & \checkmark & \checkmark & Sim \\
DMMCP\cite{cai2023consensus} & 2023 & Obj & C+L &   &   & Interm. & \checkmark & \checkmark & \checkmark & Sim+Real \\
SAPI\cite{jia2024hierarchical} & 2024 & Feat & C+L &   &   & Interm. & \checkmark & \checkmark & \checkmark & Sim \\
IRS-enhanced\cite{qi2025irs} & 2025 & Feat & C+L+R &   &   & Interm. & \checkmark & \checkmark & \checkmark & Sim \\

\rowcolor{gray!15}
\multicolumn{11}{l}{\textit{Hierarchical Federated Cooperative Architectures}} \\

Cooper\cite{chen2019cooper} & 2019 & Raw & L & \checkmark &   & Early & \checkmark &   & \checkmark & Sim \\
F-cooper\cite{chen2019f} & 2019 & Feat & L & \checkmark & \checkmark & Interm. & \checkmark &   & \checkmark & Sim \\
VI-CP\cite{mo2022method} & 2022 & Obj & C+L & \checkmark & \checkmark & Late & \checkmark &   & \checkmark & Sim+Real \\
FCP\cite{chi2023federated} & 2023 & Feat & L & \checkmark & \checkmark & Interm. & \checkmark & \checkmark & \checkmark & Sim \\
FedDWA\cite{zhang2024federated} & 2024 & Feat & L & \checkmark & \checkmark & Interm. & \checkmark & \checkmark & \checkmark & Sim \\
PEFLA\cite{chi2025parameter} & 2025 & Feat & L & \checkmark & \checkmark & Interm. & \checkmark & \checkmark & \checkmark & Sim \\

\rowcolor{gray!15}
\multicolumn{11}{l}{\textit{Hybrid-Adaptive Architectures}} \\

S-AdaFusion\cite{Qiao_2023_WACV} & 2023 & Feat & L &   & \checkmark & Interm. & \checkmark & \checkmark & \checkmark & Sim \\
ADGE\cite{Kuang2024Fast} & 2024 & Feat & L &   & \checkmark & Interm. & \checkmark & \checkmark & \checkmark & Sim \\
ACP\cite{Pillar2024} & 2024 & Feat & L &   & \checkmark & Interm. & \checkmark & \checkmark & \checkmark & Sim \\
MARL\cite{qu2024model} & 2024 & Feat & C+L & \checkmark & \checkmark & Interm. & \checkmark & \checkmark & \checkmark & Sim \\
RL-TSC\cite{LI2024104860} & 2024 & Obj & C+L & \checkmark & \checkmark & Late & \checkmark & \checkmark & \checkmark & Sim+Real \\
AccBEV\cite{shi2025v2v} & 2025 & Feat & C+L & \checkmark & \checkmark & Interm. & \checkmark & \checkmark & \checkmark & Sim \\
MDNet\cite{he2025mdnet} & 2025 & Feat & C+L & \checkmark & \checkmark & Interm. & \checkmark & \checkmark & \checkmark & Sim \\
\hline
\end{tabular}}
\vspace{0.1pt}
\noindent
\begin{tabular}{p{\dimexpr\linewidth-2\tabcolsep\relax}}
\rowcolor{yellow!10}
\footnotesize
\textbf{Note:}
\checkmark~=~Available or supported. 
\textbf{Data Granularity:} Obj—object-level, Feat—feature-level, Raw—raw data. 
\textbf{Modality:} C—camera, L—LiDAR, R—Radar.
\textbf{Central:} Indicates the presence of a centralized coordinator or cloud server for global fusion or task allocation. 
\textbf{Edge:} Represents computation or fusion performed at edge devices or RSUs (e.g., MEC nodes).
\textbf{Fusion:} Early—early fusion, Interm.—intermediate fusion, Late—late fusion. 
\textbf{Scalability:} Capability to maintain performance under increasing agents or traffic density. 
\textbf{Adaptivity:} Ability to adjust to varying topology, bandwidth, or environmental conditions. 
\textbf{Efficiency:} Communication and computational efficiency. 
\textbf{Evaluation Scope:} Sim—simulation, Real—real-world experiments, or hybrid setups.
\\
\hline
\end{tabular}
\end{table*}

The structural organization of cooperative perception systems defines how sensing, fusion, and communication modules are interconnected to achieve collective environmental awareness. 
Beyond individual perception pipelines, cooperative perception systems differ fundamentally in their architectural topologies---that is, the manner in which information is aggregated, distributed, and coordinated among vehicles, RSUs, and edge servers. 
These organizations determine not only the flow of perceptual information but also the balance among latency, scalability, and reliability. 
Four primary organizational paradigms can be identified in contemporary literature: centralized or edge-centric, decentralized or peer-to-peer, hierarchical or federated, and hybrid or adaptive structures. 
Together, they form a continuum from tightly coordinated architectures to dynamically adaptive networks, illustrating how cooperative perception evolves toward more intelligent and responsive cooperative ecosystems.

\subsubsection{Centralized Architectures}

Centralized architectures represent one of a well-established paradigms in cooperative perception, especially for intersection and corridor scenarios where occlusion and multi-agent congestion are prevalent. In such systems, vehicles and RSUs typically uplink raw sensor data or compact feature representations to an offboard processor for global fusion and scene reconstruction. While early centralized cooperative perception relied on remote cloud servers, modern deployments increasingly integrate \emph{edge computing} to reduce latency and bandwidth costs. The edge node, usually implemented as a roadside multi-access edge computing (MEC) server or an RSU cluster, acts as a localized perceptual hub—merging heterogeneous viewpoints, performing detection or tracking in a unified world coordinate system, and disseminating processed results back to vehicles as safety-ready cues. This ``edge-assisted centralization'' effectively brings the benefits of global fusion closer to data sources, enabling real-time operation even in dense or high-speed traffic. Empirical evidence from \emph{EdgeCooper}~\cite{luo2023edgecooper} and edge-based BEV frameworks~\cite{lin2024edge} shows that as the number of cooperative vehicles increases, both detection range and robustness improve, while network-aware scheduling mechanisms maintain accuracy under bandwidth constraints. These results demonstrate that edge-level coordination achieves a favorable balance among spatial coverage, latency, and communication cost, forming the practical backbone of contemporary centralized cooperative perception deployments.

From a deployment perspective, the roadside layer serves as the structural backbone of centralized cooperative perception. Elevated RSU viewpoints enable persistent cross-lane and intersection monitoring, allowing centralized fusion to overcome line-of-sight limitations and local sensor blind spots. The large-scale \emph{RCooper} dataset~\cite{hao2024rcooper} illustrates this principle by integrating roadside cameras and LiDARs across intersections and corridors to produce globally consistent tracklets for vehicle consumption. Complementary research on LiDAR placement optimization~\cite{lin2025roadside} further highlights that strategic RSU deployment significantly enhances detection recall under dynamic occlusion patterns. Conceptually, centralized and edge-centric architectures embody the transition from \emph{shared sensing} to \emph{shared perception}: they structure multi-agent evidence at the edge and generate calibrated, time-stamped, regionally consistent world models. Within the broader SPD framework, such architectures bridge the sensing and perception layers, ensuring that downstream decision-making modules receive reliable, synchronized environmental understanding necessary for cooperative safety intelligence~\cite{yu2022review,chen2024smart}.

\subsubsection{Decentralized Architectures}
Decentralized architectures organize cooperation directly among vehicles through V2V broadcasts, with each agent running its own perception, maintaining local tracks, and exchanging compact information (e.g., object or track states) to enlarge awareness and improve consistency without a central coordinator. Object/track–level exchange aligns well with standardized BSM/CAM/CPM messages and supports multi-sensor AV stacks; recent consensus-based distributed fusion shows how local sensor fusion and connected-node fusion can be unified so that vehicles collectively improve detection and tracking of non-connected targets while keeping communication loads moderate (including nonlinear CKF-based updates and multi-model handling for maneuvering targets)~\cite{cai2023consensus}. System-level evaluations of V2V cooperative perception frameworks report favorable perception accuracy and coverage across urban and highway scenarios, and characterize scalability with respect to participation rates and ad-hoc losses, highlighting that decentralized data association and fusion can serve large, mixed-traffic environments when broadcast reliability and participation are considered in the design of the metric and pipeline~\cite{yoon2021performance}. Complementary studies of decentralized resource allocation emphasize relevance-aware scheduling in congested channels: vehicles with more critical intent or information claim higher message rates while others yield, improving application-layer delivery without sacrificing radio performance---an approach that fits P2P cooperative perception where agents negotiate bandwidth collaboratively at the edge~\cite{bischoff2021prioritizing}. At the timing layer, overviews focused on time-sensitive cooperative perception frame P2P cooperation as one of two primary classes (alongside infrastructure-assisted IoT or smart-city cooperation), outlining delay budgets, synchronization, congestion control, and fusion choices needed to sustain real-time perception among moving peers~\cite{aoki2022time}.

Beyond pure perception fusion, decentralized cooperative perception increasingly incorporates graph and attention mechanisms and hierarchical coordination to adapt to dynamic topology: distributed consensus and weighting schemes help align estimates among neighbors under heterogeneous sensors and motion models~\cite{cai2023consensus}, while hierarchical perception-improving frameworks from decentralized multi-robot navigation demonstrate sensor-wise and agent-wise attention to aggregate features from arbitrary neighbor sets---an idea transferable to road traffic for scalable, mapless cooperation under limited communications~\cite{jia2024hierarchical}. Emerging wireless designs also point to P2P-friendly enhancements: joint broadcast and multicast optimization with programmable surfaces targets ultra-reliable low-latency delivery of perception payloads and task-aware resource mapping in high-mobility groups---promising ingredients for robust P2P cooperative perception on 5G and 6G V2X as groups self-organize without fixed coordinators~\cite{qi2025irs}. Taken together, these threads portray decentralized cooperative perception as a flexible, resilient organization: vehicles keep autonomy of perception and decision, exchange semantically meaningful summaries, prioritize relevant information, and adapt communication and fusion to topology and load---properties that align naturally with the SPD trajectory from shared sensing toward shared, actionable understanding.

\subsubsection{Hierarchical Federated Cooperative Architectures}
Hierarchical (vehicle--edge/RSU--cloud) federated organizations bridge localized perception with global situational understanding by distributing sensing, fusion, and learning across multiple tiers. Within such multi-tier systems, low-latency sensing and preliminary fusion are performed on vehicles, regional aggregation and time-critical processing occur at RSUs or edge servers, and long-horizon reasoning and cross-region optimization are handled in the cloud. Early implementations established this structure in practice: \emph{Cooper} demonstrated raw point-cloud aggregation across vehicles to extend sensing coverage, while explicitly addressing bandwidth and pose alignment constraints that motivate offboard fusion. \emph{F-Cooper} advanced this idea through an edge-assisted variant where vehicles transmit compact LiDAR features to an edge server for real-time fusion, achieving significant perception gains under strict latency budgets---an archetypal ``vehicle $\rightarrow$ edge $\rightarrow$ feedback'' loop for complex intersections and occluded urban zones. Collectively, these works highlight how hierarchical placement of fusion modules can balance computational load, semantic abstraction, and communication cost~\cite{chen2019cooper,chen2019f}.

As cooperative perception matured, hierarchical architectures increasingly incorporated \emph{learning hierarchy} to align local adaptation with global consistency. Federated cooperative perception has emerged as a natural evolution, enabling privacy-preserving, bandwidth-efficient collaboration across diverse agents. Recent frameworks introduce dynamic client weighting and distribution-aware losses to mitigate non-IID effects and sensor heterogeneity (e.g., cars, buses, and trucks), thereby improving BEV segmentation performance on federated datasets~\cite{zhang2024federated}. Meanwhile, vehicle--road--cloud cooperation studies emphasize the complementary roles of each tier: RSUs maintain persistent regional context, edge nodes ensure temporal continuity, and cloud services provide long-term supervision and policy refinement~\cite{yu2022review,gao2024vehicle}. At the algorithmic level, hierarchical Kalman-style fusion further validates the reliability of such multi-layer coordination, showing that adaptive gain switching can preserve estimation stability even under partial RSU dropouts~\cite{mo2022method}. Taken together, hierarchical and federated frameworks partition perception and learning according to timing and scope—vehicles for immediacy, edges for consensus, and clouds for knowledge retention—creating a scalable, privacy-aware, and resilient foundation for cooperative perception in connected transportation ecosystems.

\subsubsection{Hybrid-Adaptive Architectures}
Hybrid-adaptive organizations integrate the strengths of centralized, decentralized, and hierarchical cooperative perception, while allowing the cooperation mode to vary with bandwidth, latency, task urgency, and environmental context. A representative line of work treats \emph{communication as a controllable resource} within the perception loop~\cite{LI2024104860,shi2025v2v}. For instance, the Pillar Attention Encoder (PAE) enables \emph{adaptive cooperative perception} by learning the significance of LiDAR pillar features and filtering what to transmit under dynamic capacity, thereby relaxing the common assumption that each node can share identically sized features~\cite{Pillar2024}. PAE introduces a pillar-level attention mechanism and an \emph{adaptive feature filter} that retains salient features for sharing; evaluations vary the transferable ratio $\omega$ (without re-training across bandwidth regimes) to examine robustness under communication loads~\cite{he2025mdnet}. Together, these mechanisms enable efficient feature fusion despite heterogeneous, time-varying link budgets and participant sets, providing a practical path for content-aware scaling in real deployments~\cite{qu2024model}.

In a broader sense, adaptivity extends from feature filtering to \emph{cooperation scheduling} and resource management. Model-assisted multi-agent reinforcement learning frameworks dynamically switch between stand-alone and cooperative perception on a per-pair basis, while jointly allocating radio and computing resources. This task-aware control maximizes perception efficiency under delay constraints and mobility-driven channel variations---a hallmark of self-optimizing cooperative intelligence~\cite{qu2024model}. At the traffic-system scale, cooperative perception has also been embedded into adaptive signal control: by fusing cooperative perception-enabled observations with a cell-transmission traffic model (\emph{CAVLight}), deep reinforcement learning policies reduce delay and remain effective even at low CAV penetration, illustrating cross-layer benefits when perception and control are co-designed~\cite{LI2024104860}. Meanwhile, robustness against \emph{communication uncertainty} has been explored through adaptive lossy-channel modeling for multi-vehicle BEV perception (\emph{AccBEV}). A variational framework reconstructs shared BEV features while an adaptive channel module compensates for degradation across SNR states, sustaining cooperation under fluctuating network conditions~\cite{shi2025v2v}. 

Taken together, hybrid-adaptive cooperative perception frameworks form a closed perception--communication--action loop that aligns feature selection, cooperation scheduling, resource allocation, and decision support under dynamic deployment constraints. This pattern reflects a shift from fixed data exchange toward context-sensitive coordination, making adaptive cooperation an important bridge between perception-layer evidence and decision-layer safety intelligence.

\subsubsection{Comparative Insights and Systemic Implications}

The comparative analysis of centralized, decentralized, hierarchical, and hybrid organizations reveals both the technological maturity and intrinsic constraints of current cooperative perception paradigms. Each structure prioritizes one aspect of the perception--communication--computation triad but falls short of achieving full systemic intelligence. Centralized or edge-centric architectures excel in producing globally consistent situational awareness, yet they depend heavily on stable connectivity and infrastructure density, making them vulnerable to latency spikes or network failures. Decentralized and peer-to-peer frameworks, while robust to disconnections, struggle with consistency, redundancy control, and the absence of a unified semantic frame. Hierarchical and federated systems introduce scalability and multi-level optimization but often face synchronization bottlenecks between tiers and slow feedback propagation. Even hybrid and adaptive designs—though capable of dynamic mode switching—typically treat communication and perception as separable optimization targets, lacking a unified notion of reasoning or intent alignment across agents.

These systemic limitations highlight the transition point from cooperative perception to cooperative intelligence. The evolution toward the unified SPD framework reflects a paradigm shift from passive data sharing to active, feedback-driven reasoning. SPD couples perception with decision-making, embedding adaptive guidance, priority scheduling, and bidirectional calibration within the perception process itself. By aligning perception with decision-oriented safety goals, SPD helps what is perceived, shared, and inferred collectively support higher-level safety objectives under PQoS constraints. In this view, cooperative perception becomes the sensory substrate of an embodied reasoning loop that integrates observation, cognition, and intervention to support proactive safety intelligence in V2X ecosystems.

\section{Cooperative Safety Intelligence along the SPD Framework}

\begin{figure}[t]
  \centering
  \includegraphics[width=0.9\linewidth]{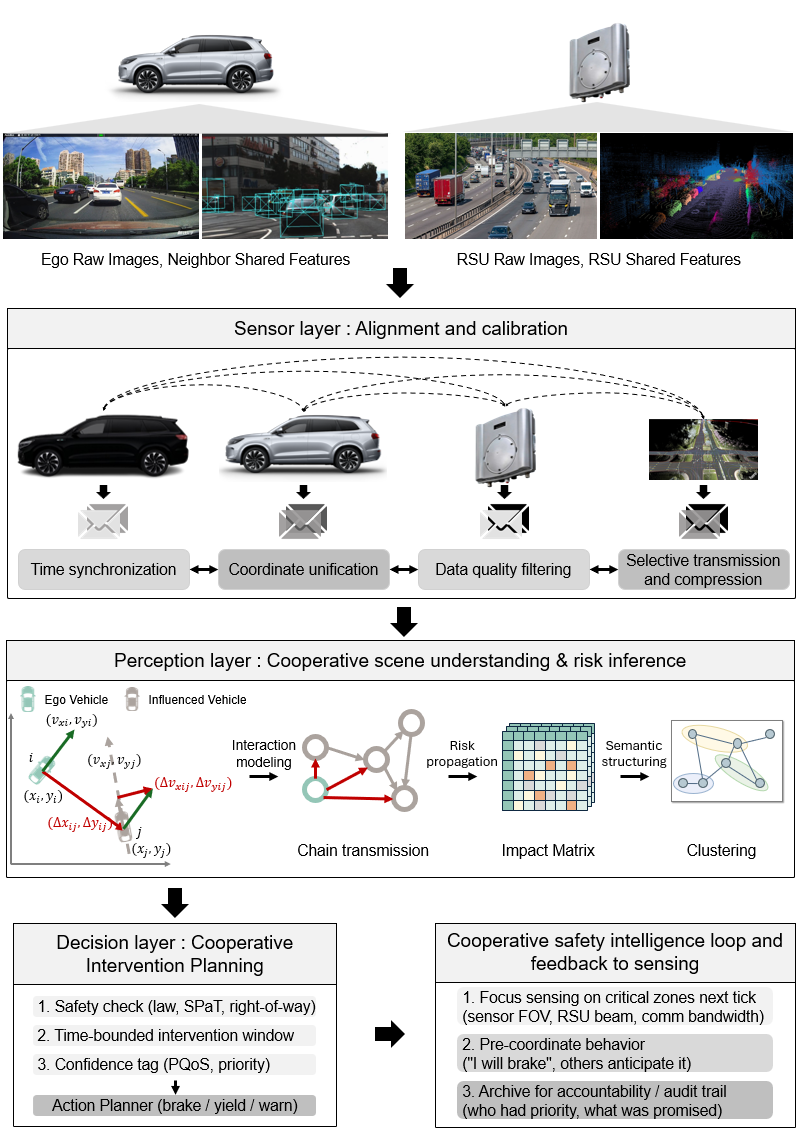}
  \caption{Unified \textbf{SPD} framework for cooperative safety intelligence. 
  The framework depicts how sensing provides structured evidence, perception transforms evidence into predictive understanding, 
  and decision feeds back into sensing through adaptive guidance and calibration.}
  \label{fig:spd_framework}
\vspace{-6mm}
\end{figure}

This section operationalizes cooperative intelligence through the unified 
SPD framework, as depicted in Fig.~\ref{fig:spd_framework}.
The framework defines how sensing provides structured evidence, perception 
transforms evidence into predictive understanding, and decision feeds back 
into sensing through adaptive guidance and calibration.

\begin{itemize}
    \item \textbf{Sensor layer.} The sensor layer of the SPD framework provides the physical and communication foundations of cooperative safety intelligence. 
    It explains how distributed sensing assets, communication links, timing bases, standardized message primitives, and deployment architectures jointly determine what a system can \emph{see} and how reliably that information propagates to the next stage.

    \item \textbf{Perception layer.} The perception layer of the SPD framework transforms distributed sensor signals into structured, trustworthy, and forecast-ready evidence. 
    It explains how cooperative perception, multimodal fusion, and synchronization mechanisms jointly ensure that safety systems not only \emph{see} but also \emph{understand} and \emph{trust} what they see.

    \item \textbf{Decision layer.} The decision layer translates calibrated, time-stamped evidence from perception into timely, coordinated, and human-aligned actions. 
    It describes how risk evolves from probabilistic forecasts to concrete triggers and shared maneuvers across agents and humans.
\end{itemize}

\subsection{Sensor Layer for Safety: V2X Foundations and Evidence Acquisition}

The section is organized into four subsections: (1) \emph{BLOS}, detailing how RSUs, vehicles, and UAVs extend visibility beyond line of sight; (2) \emph{V2X Communication Models}, outlining V2V, V2I, V2P, and V2N and their safety roles; (3) \emph{Time \& Positioning Bases (GNSS/Network Time)}, describing shared spatio-temporal references for synchronized evidence; (4) \emph{Message Primitives and Data Carriers}, explaining how standardized data units and link technologies turn physical observations into structured evidence; and (5) \emph{Deployment \& Edge (Density, MEC, Relays, UAVs)}, showing how placement, edge computing, and relays operationalize coverage and latency at scale. Together, these layers define the technical substrate upon which perception and forecasting operate, linking sensing fidelity to downstream safety readiness.

\subsubsection{Sensing Assets and Coverage}

Within the SPD framework, the sensor layer functions as the evidential foundation of cooperative safety intelligence. 
It \emph{structures, synchronizes, and qualifies} sensory observations, enabling subsequent perception and decision layers to operate on consistent, trustworthy information.
Through distributed V2X infrastructures—vehicles, RSUs, and UAVs—the sensor layer transforms raw observations into \emph{safety-ready evidence}: observations annotated with temporal validity, spatial alignment, confidence, and communication latency. 
This evidence becomes the input substrate for perception-level fusion and forecasting, while its timing and reliability metadata directly inform decision-level reasoning and intervention timing. 

Perception for safety begins with a simple but demanding requirement: the system is required to perceive safety-critical elements with sufficient lead time \cite{delooz2022analysis}. Ego sensors offer the baseline, yet line-of-sight and urban geometry constrain how far and how stably they can ``see,'' especially at occluded locations—such as a pedestrian stepping from behind a van, cross-traffic masked by a truck, or a merge hidden by road curvature \cite{xu2024v2x,mouawad2021collective}. Extending visibility is therefore a primary and tangible contribution of V2X-enabled perception. By pooling vantage points across vehicles and infrastructure, cooperative systems can deliver an expanded horizon beyond what ego perception alone typically attains \cite{bazzi2020wireless}.

Early demonstrations relied on direct V2V broadcasts, where nearby cars share positions and maneuvers at low latency to warn of blind-spot hazards~\cite{eldeeb2023energy}. This mode remains essential because it operates without fixed infrastructure and issues urgent alerts within milliseconds \cite{ngo2023cooperative}. In complex intersections and merging zones, infrastructure becomes a decisive complement. RSUs mounted at elevated viewpoints provide continuous, persistent coverage decoupled from individual vehicle motion \cite{ko2020rsu}. Recent road-to-vehicle vision treats RSUs as long-lived sensors: they aggregate scene dynamics, maintain temporal ``slices'' of interactions, and deliver compact, context-rich features to approaching vehicles~\cite{tan2024dynamic}. By supplying continuity in space and time, RSUs strengthen range and persistence beyond ego-only sensing \cite{arnold2020cooperative}.

\begin{figure}[t]
    \centering
    \includegraphics[width=0.9\linewidth]{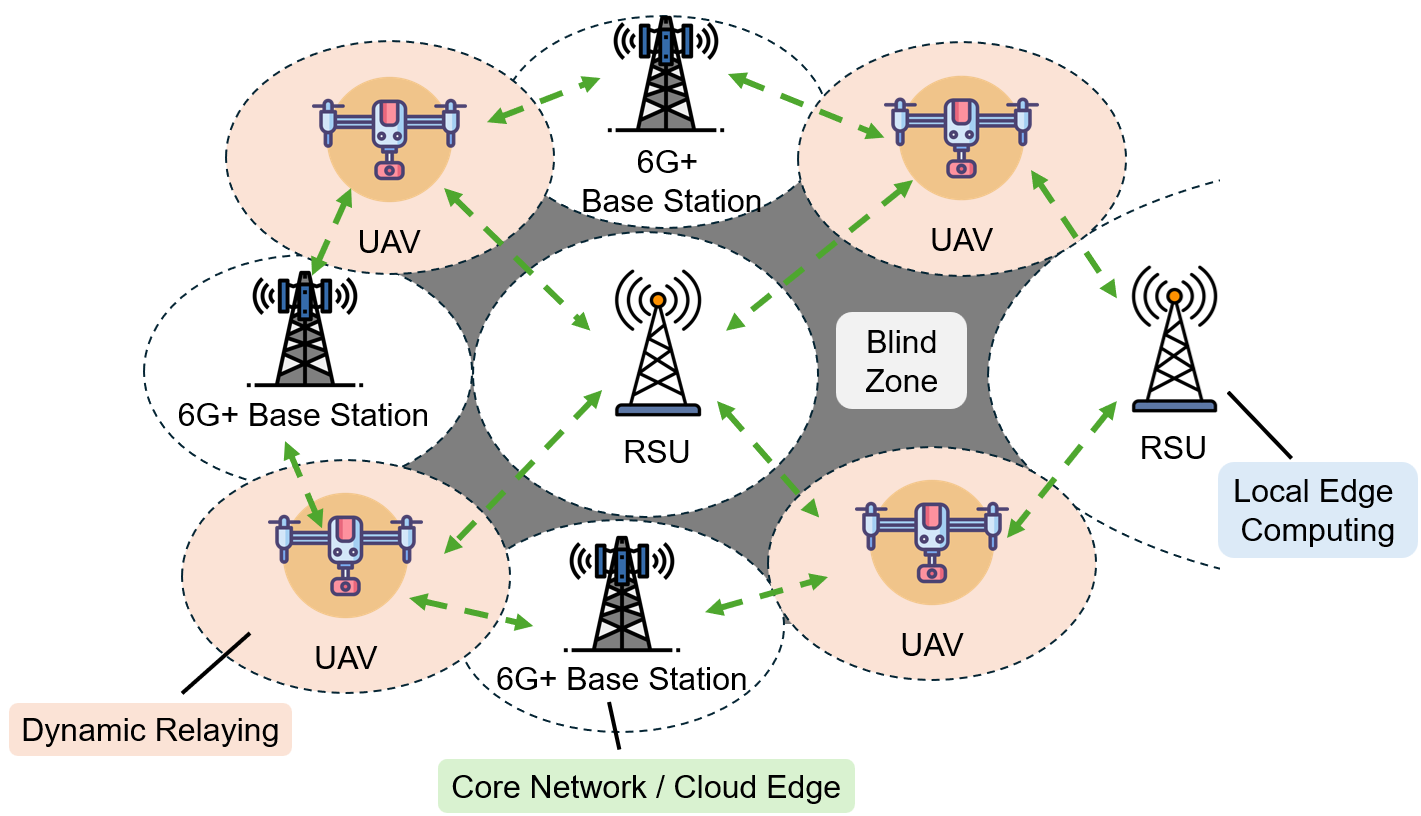}
    \caption{UAV-assisted RSUs for closing coverage ``blind zones.'' RSUs provide stable access points and local edge computing, while UAVs act as dynamic relays to 
    extend connectivity where roadside infrastructure is insufficient. Together with 6G+ base 
    stations and cloud/edge backends, this aerial–terrestrial system sustains V2X services in areas 
    otherwise left uncovered~\cite{andreou2023uav}.}
    \label{fig:rsu}
    \vspace{-6mm}
\end{figure}

Fixed infrastructure pairs naturally with agile complements. UAVs relays provide flexible extensions to ground RSUs \cite{demir2020energy}. Aerial vantage points mitigate coverage gaps in suburban corridors or temporary work zones, and stochastic-geometry placement (e.g., Voronoi–Poisson) positions UAVs to maximize effective visibility with minimal overhead~\cite{andreou2023uav}. As Fig.~\ref{fig:rsu} illustrates, RSUs anchor stable access and edge compute, while UAVs extend the footprint into previously unserved regions. This aerial–terrestrial pairing delivers resilience that static deployments alone seldom achieve.

\begin{figure}[t]
    \centering
    \includegraphics[width=1\linewidth]{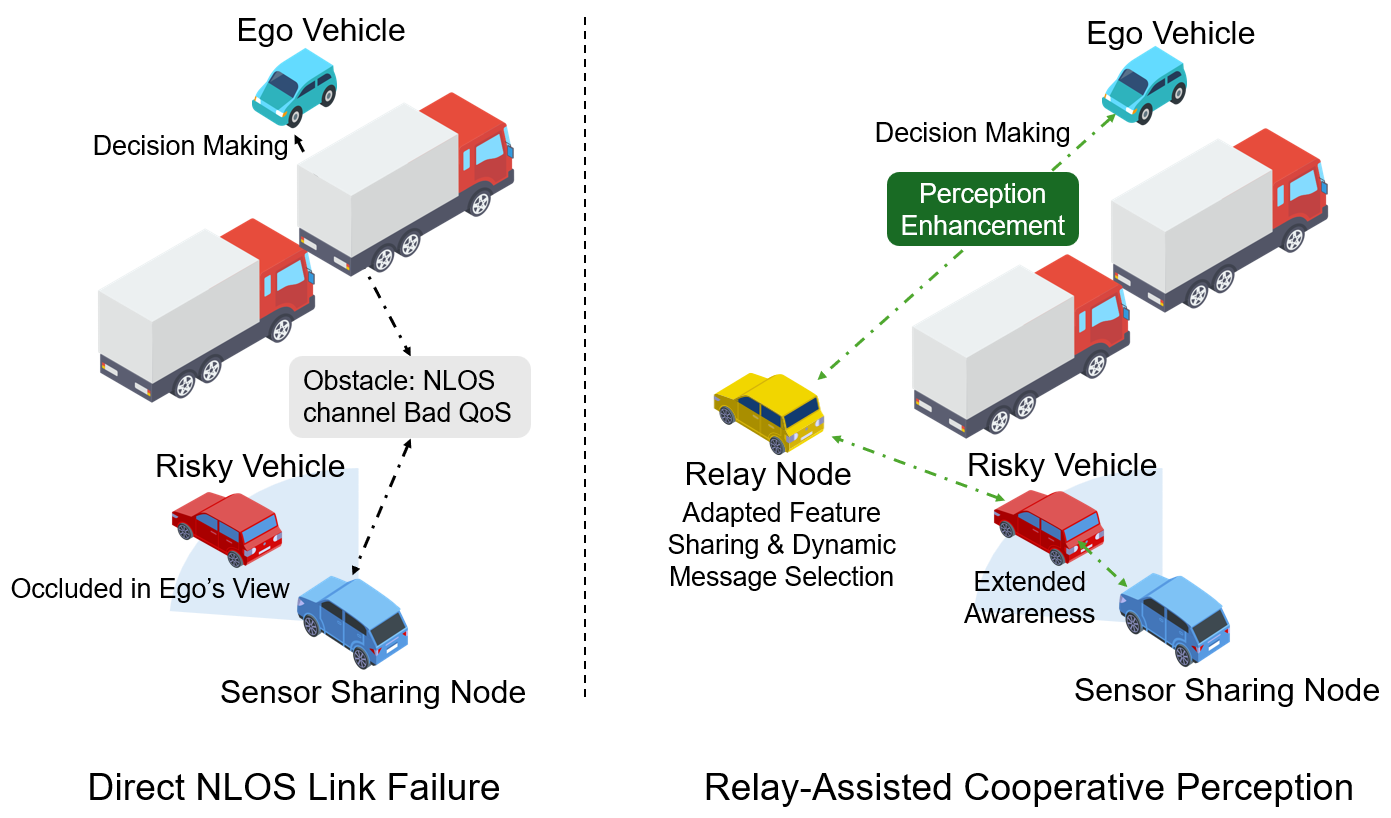}
    \caption{Relay-assisted cooperative perception under NLOS conditions. 
    When direct V2V links are blocked by large obstacles (left), nearby relay nodes can sustain feature-level sharing and maintain perception quality (right).\cite{li2024nlos}}
    \label{fig:nlos}
    \vspace{-5mm}
\end{figure}

Visibility is simultaneously a geometric and a channel challenge. NLOS occlusions attenuate optical/radar returns and sidelink signals \cite{li2024nlos}. The same truck that hides a crossing car may also weaken the cooperative message intended to reveal it. Robust designs therefore integrate perception and networking. Relay selection and store–carry–forward sustain feature-level sharing under NLOS, and empirical studies show that multi-hop paths preserve perception quality in obstructed corridors (Fig.~\ref{fig:nlos})~\cite{li2024nlos}. Opportunistic relays—neighboring cars or UAVs—thus act as visibility enablers as much as communication nodes \cite{mao2024high}.

Large-scale deployments have validated the significance of extended cooperative sensing. In the New York City Connected Vehicle Pilot Deployment (NYC CVPD), approximately 450 RSUs and about 3,000 equipped city vehicles supported V2V, V2I, and pedestrian-oriented safety applications, including BLOS-related warnings, SPaT/MAP-based intersection advisories, and pedestrian safety services~\cite{yang2022functional}.
Complementary 5G NR-V2X prototypes equipped with MEC further demonstrate that low-latency retransmission and edge offloading can sustain perception and safety services even under heavy network load~\cite{lucas2022analytical}. Collectively, these deployments illustrate how extended visibility can be realized and maintained at city scale when technology design, communication reliability, and policy coordination are jointly aligned. From the perspective of the SPD framework, such cooperative sensing configurations define the \emph{observable horizon} of the safety loop—determining how early potential risks can be detected, structured into shareable evidence, and propagated to perception and decision layers for timely reasoning and intervention.

From a safety perspective, visibility beyond line of sight enriches predictive evidence. With extended views, forecasting modules draw on spatially and temporally broader context, enabling the forecast–risk loop to anticipate conflicts earlier and more reliably. The next requirement is to deliver this visibility in time and in sync with the traffic context.

\subsubsection{Time \& Positioning Bases (GNSS/Network Time)}

Accurate sensing and cooperative awareness depend on a unified spatio-temporal reference. GNSS provides absolute positioning, while urban canyons and tunnels are mitigated through cooperative localization that fuses GNSS with relative ranging and V2X-based measurements.
For example, CV2X-LOCA achieves lane-level accuracy using RSU channel-state information under GNSS-challenged conditions~\cite{huang2024toward}, while implicit and DNN-assisted methods aggregate LiDAR or range cues across vehicles to reduce positional uncertainty~\cite{qu2023cooperative}.
Temporal alignment is maintained through network-distributed clocks: 5G NR V2X Release~16 introduces the Network Time Distribution Service (NTDS), achieving microsecond-level synchronization via GNSS time or IEEE~1588 PTP~\cite{garcia2021tutorial}.
Upcoming releases further integrate joint communication and sensing (JCS), allowing a single waveform to support both data exchange and radar-like perception~\cite{decarli2024performance}.
Such unified time bases ensure that cooperative messages, SPaT/MAP updates, and forecasting windows remain temporally consistent, while precise time and position tags underpin PQoS metadata and evidence contracts used across the SPD loop.

The evolution toward \emph{Transportation~5.0} brings a paradigm shift from centralized synchronization to service-oriented, decentralized coordination.
The Decentralized Autonomous Service (DAS) framework~\cite{gao2024cooperative} reimagines localization and timing as distributed micro-services shared across vehicles, RSUs, and edge nodes.
Each agent both contributes to and consumes calibrated references, forming a resilient network that self-corrects under partial outages and variable connectivity.
DAS thus extends GNSS and network time into an autonomous infrastructure layer—maintaining consistent spatio-temporal awareness for cooperative forecasting, risk evaluation, and intervention planning.
From the SPD perspective, these shared time and positioning foundations ensure that cooperative evidence remains chronologically and spatially coherent, enabling perception modules to fuse multi-agent observations on a consistent timeline and decision modules to act upon synchronized, trustworthy situational awareness.

After establishing time and positioning references, cooperative safety systems standardize \emph{what} information is exchanged and \emph{how} it is conveyed. Message primitives define the semantics of safety-relevant data, turning physical observations into structured digital evidence intelligible to heterogeneous agents. They specify the type, frequency, and scope of information broadcast by each node—vehicle, infrastructure, or pedestrian device—forming the linguistic foundation of vehicular cooperation. Standards families (SAE~J2735, ETSI~EN~302~637, ISO~19091) align these semantics across regions \cite{yoshizawa2023survey}.

Core units such as the Basic Safety Message (BSM) and Cooperative Awareness Message (CAM) report dynamic states including position, velocity, heading, and maneuver intent. The Decentralized Environmental Notification Message (DENM) adds event-triggered alerts, while SPaT and MAP provide intersection geometry and signal status. The Roadside Safety Message (RSM) contributes infrastructure-perceived objects and trajectories, bridging perception and communication.

Broadcast efficiency remains pivotal for wide-area safety dissemination. MBMS/eMBMS/NR-MBS enable one-to-many delivery without saturating uplink bandwidth. Geo-casting and localized MBMS align dissemination with relevant regions, while careful core-network routing preserves timeliness~\cite{tan2024dynamic,liu2022energy}. These carrier-layer refinements directly shape the completeness and latency of message primitives delivered to perception and forecasting modules, strengthening how cooperative systems anticipate and react to evolving risks. Standardized message semantics thus serve as the language through which the sensor layer communicates structured evidence to the perception and decision layers within the SPD framework.

\subsubsection{Trends and Integration}
Emerging 5G/6G systems blur the boundary between communication and sensing. Physical waveforms increasingly serve dual roles in JCS frameworks~\cite{zhong2022empowering,decarli2024performance}. Time-synchronized feature sharing lets vehicles and RSUs transmit not only message primitives but also preprocessed perception embeddings with PQoS tags. As carriers evolve from packet transport to semantic conduits, the cooperative safety stack advances from \emph{message passing} to \emph{evidence fusion}, ensuring that perception, forecasting, and decision operate over temporally consistent, context-rich streams. The next section examines how synchronized carriers and structured messages converge to form the perception backbone of safety-ready intelligence systems.

\subsubsection{Deployment \& Edge (Density, MEC, Relays, UAVs)}

Onboard sensors offer high-fidelity local context, and complementary infrastructure extends that context beyond occlusions and short planning horizons \cite{giuliano2023musli}. RSUs provide stable viewpoints and persistent presence at conflict points, while edge nodes aggregate multi-source observations into compact products for real-time consumption by forecasting and risk modules. Recent RSU-centric systems make this concrete: adaptive road-to-vehicle vision treats the RSU as a long-lived sensor and hub that fuses multi-agent features and retains ``scene slices'' to bridge cross-scene shifts; features can be aggressively compressed (e.g., $32\times$) with marginal loss in segmentation utility, keeping bandwidth within link budgets \cite{tan2024dynamic}. Beyond perception, RSU signals themselves can serve as state estimates: an RSU-based C-V2X localization framework (CV2X-LOCA) achieves lane-level accuracy in GNSS-challenged urban canyons using channel-state information, combining coarse positioning, environment-parameter correction via cooperative RSUs, and trajectory filtering (e.g., Unscented Kalman Filter). Field tests show robust performance even with sparse RSU spacing, offering practical guidance on cost-effective deployments \cite{huang2024toward,ko2021v2x}.

When coverage varies, store–carry–forward and density-aware relaying maintain essential cues \cite{souri2024systematic}. Where roadside installations leave ``blind zones,'' UAVs dynamically extend coverage to uncovered areas, reducing perception gaps without requiring dense fixed infrastructure \cite{andreou2023uav, fernando2024uav}. Small engineering choices also compound gains: antenna placement and vehicle-body shadowing materially influence effective SNR; fast surrogate electromagnetic models shorten the iterate–measure loop, guiding targeted improvements where forecasting and warning benefit \cite{zhao2019fast,chetlur2019coverage}. By integrating edge intelligence and adaptive deployment, the sensor layer closes the loop between sensing and decision—enabling cooperative systems to actively manage visibility, latency, and evidence quality according to safety priorities.

\subsection{Perception Layer for Safety: Cooperative Evidence, Timeliness, and Forecast-Ready Outputs}
\noindent

The perception layer bridges the sensing foundation and the decision logic within the SPD framework. 
Its primary task is to transform heterogeneous observations from distributed agents into 
\emph{structured, temporally aligned, and uncertainty-aware evidence} that can be directly consumed by forecasting and risk reasoning modules. 
Rather than treating perception as an isolated visual process, this layer functions as a cooperative reasoning hub that consolidates what multiple agents observe, ensures when these observations are valid, and quantifies how reliable they are. 
It embodies the transition from observation to anticipation—turning raw sensor feeds into calibrated, context-rich evidence streams.

\subsubsection{From Cooperative Inputs to Structured Evidence}

The perception layer begins by transforming raw, heterogeneous sensor observations into structured and shareable evidence. 
Cooperative perception extends an agent’s awareness beyond ego sensors by combining inputs from multiple vehicles, RSUs, and occasionally UAVs. 
Depending on bandwidth, latency budgets, and safety criticality, 
the content exchanged between agents can take three progressively abstract forms:

\begin{itemize}
    \item \textbf{Raw-level sharing} transmits unprocessed sensor signals such as LiDAR point clouds or camera frames, 
    offering maximal fidelity but imposing high communication and processing costs~\cite{howe2021weakly}.
    \item \textbf{Feature-level sharing} transmits intermediate feature maps or embeddings extracted by deep encoders. 
    This representation preserves geometric and semantic information essential for forecasting while remaining efficient and compatible with NR-V2X link capacity~\cite{clancy2024wireless,garcia2021tutorial}.
    \item \textbf{Object-level sharing} delivers structured entities such as tracked objects, occupancy grids, or intent tokens directly aligned with standardized V2X formats (BSM, MAP, SPaT, RSM). 
    These messages are readily consumable by downstream reasoning and planning modules~\cite{kiela2020review}.
\end{itemize}

In practice, the appropriate level of sharing behaves as a dynamic policy that adapts to environmental and network conditions. 
Under high load or limited bandwidth, perception nodes may compress, sparsify, or switch from raw to feature or object sharing while maintaining core semantics~\cite{tan2024dynamic}. 
Adaptive encoding and bandwidth-aware prioritization ensure that safety-critical agents (e.g., vehicles on collision courses) retain high fidelity, 
whereas less relevant context is represented more compactly~\cite{hu2024multi,mlika2022deep}. 
This flexibility keeps evidence both actionable and sustainable under real-world constraints.

Occlusions and NLOS conditions further complicate this process. 
The same physical obstacle that blocks a camera’s view may also attenuate V2V communication, reducing QoS and threatening evidence continuity. 
Relay-assisted cooperative perception resolves this bottleneck by introducing opportunistic relay nodes—neighboring vehicles or UAVs—that re-route or regenerate feature-level messages to restore line-of-sight connectivity (Fig.~\ref{fig:nlos})~\cite{li2024nlos}. 
Recent frameworks integrate link-state awareness directly into perception modules, 
using transformer-based fusion to co-optimize sensing and communication: the system learns when to trigger relays, how to allocate bandwidth, and which modalities to prioritize~\cite{gharsallah2024mvx,liu2019edge}. 
Such cross-domain designs ensure that perception retains geometric integrity and temporal continuity even in congested or obstructed urban canyons.

From the SPD perspective, this cooperative evidence formation represents the first transformation of sensing outputs into structured, trust-calibrated knowledge. 
By hierarchically abstracting observations (raw $\rightarrow$ feature $\rightarrow$ object) and adapting their granularity to link conditions, 
the perception layer constructs reliable, communication-aware evidence that downstream forecasting and decision modules can directly interpret as the foundation of safety reasoning.

\subsubsection{Multi-Modal and Multi-Agent Fusion with Benchmark Support}

\begin{table*}[t]
\caption{Datasets categorized by their \textbf{benchmark tasks, sensing setups, and cooperation level}.}
\label{tab:dataset_comparison}
\centering
\large
\setlength{\tabcolsep}{6pt}
\renewcommand{\arraystretch}{1.3}

\rowcolors{2}{yellow!10}{yellow!3} 
\resizebox{\textwidth}{!}{%
\begin{tabular}{l c c c p{11cm}}
\rowcolor{yellow!25}
\hline
\textbf{Dataset} & \textbf{Scenario} & \textbf{Source} & \textbf{Sensors} & \textbf{Tasks} \\
\hline

\rowcolor{gray!15}
\multicolumn{5}{l}{\textit{Non-cooperative baselines: Single-vehicle perception/motion forecasting benchmarks}} \\

KITTI~\cite{geiger2012we}              & single & real-world  & LiDAR      & Detection, Tracking, Segmentation, Motion (derivable) \\
nuScenes~\cite{caesar2020nuscenes}     & single & real-world  & MTV, LiDAR & Detection, Tracking, Segmentation, Motion (derivable) \\
Waymo~\cite{sun2020scalability}        & single & real-world  & MTV, LiDAR & Detection, Tracking, Segmentation, Motion (derivable) \\
Argoverse~1~\cite{chang2019argoverse}  & single & real-world  & MTV, LiDAR & 3D Tracking (113 scenes), Motion Forecasting, HD map-based planning support \\
Argoverse~2~\cite{wilson2023argoverse} & single & real-world  & MTV, LiDAR & Sensor Perception, Motion Forecasting, Map Change Analysis \\

\rowcolor{gray!15}
\multicolumn{5}{l}{\textit{Non-cooperative baselines: Single-vehicle accident anticipation / detection benchmarks}} \\

DAD~\cite{chan2017anticipating}        & single & real-world & MTV                 & {Accident anticipation} (dashcam sequences near collision) \\
CCD~\cite{bao2020uncertainty}          & single & web video  & MTV                 & {Accident anticipation} with context tags (weather/ego/cause) \\
A3D~\cite{yao2019unsupervised}         & single & real-world & MTV                 & {Accident detection \& anticipation} (dashcam) \\
DADA-2000~\cite{fang2021dada}          & single & real-world & MTV, Gaze, Semantics & {Driver attention} \& {accident anticipation} \\
VIENA2~\cite{aliakbarian2018viena}     & single & simulator  & MTV, Vehicle states  & {Driving action recognition} \& {accident anticipation} \\
GTACrash~\cite{kim2019crash}           & single & simulator  & MTV, Vehicle states  & {Danger/accident categorization} (synthetic) \\
YouTubeCrash~\cite{kim2019crash}       & single & web video  & MTV                 & {Dangerous vehicle classification} / incident flags \\
TAD~\cite{xu2024tad}                   & single & traffic CCTV & MTV                & {Traffic accident detection} \& localization (fixed views) \\

\rowcolor{gray!15}
\multicolumn{5}{l}{\textit{Cooperative perception and V2X-enabled datasets}} \\

OPV2V~\cite{xu2022opv2v}               & V2V  & simulator   & LiDAR               & \textbf{Cooperative 3D perception}: detection/tracking in BEV; \textbf{feature/object sharing} under dynamic multi-vehicle topology \\
V2X-Sim~\cite{li2022v2x}               & V2V \& V2I & simulator & MTV, LiDAR         & \textbf{Multi-agent perception} (detection/segmentation), \textbf{motion forecasting} (derivable), \textbf{communication-aware} evaluation \\
DAIR-V2X~\cite{yu2022dair}             & V2I  & real-world  & MTV, LiDAR          & \textbf{Cross-view} vehicle–RSU \textbf{detection/tracking}; real-world cooperative perception benchmark \\
V2X-Seq/Perception~\cite{yu2023v2x}    & V2I  & real-world  & MTV                 & \textbf{Synchronized multi-view perception}, temporal consistency, cross-agent calibration \\
V2X-Seq/Forecasting~\cite{yu2023v2x}   & V2I  & real-world  & MTV                 & \textbf{Cooperative motion forecasting}: aligned ego–RSU streams, long-horizon multi-agent prediction \\
DeepAccident~\cite{wang2024deepaccident} & V2V \& V2I & simulator & MTV, LiDAR        & \textbf{End-to-end cooperative safety}: detection $\rightarrow$ tracking $\rightarrow$ forecasting $\rightarrow$ \textbf{accident risk inference} with V2X fusion \\
V2XSet~\cite{xu2022v2x} & V2X & simulator & LiDAR, Camera & \textbf{Cooperative 3D detection} with V2V/V2I fusion; benchmark for intermediate-fusion perception \\
DOLPHINS~\cite{mao2022dolphins} & V2X & simulator & LiDAR, Camera & \textbf{Collaborative 2D/3D detection} across vehicles and RSUs; supports multi-level fusion \\
{V2X-R~\cite{huang2025v2x}} & V2X & simulator & LiDAR, Camera, Radar & \textbf{LiDAR–Radar 3D detection} under adverse weather; benchmark for robust fusion \\
V2XPnP-Seq~\cite{zhou2025v2xpnp} & V2X & real-world & LiDAR, Camera & \textbf{Cooperative perception \& prediction}: multi-frame Transformer fusion with HD-map context \\
SCOPE~\cite{gamerdinger2024scope} & V2X & simulator & LiDAR, Camera & \textbf{2D/3D detection} and \textbf{semantic segmentation} under physical weather; \textbf{Sim2Real} benchmark \\
\hline
\end{tabular}%
}
\vspace{0.1pt}
\noindent
\begin{tabular}{p{\dimexpr\textwidth-2\tabcolsep\relax}}
\rowcolor{yellow!10}
\footnotesize
\textbf{Note:} MTV~=~Multi-view cameras. This table integrates single-vehicle and cooperative datasets, grouped into
(1) perception/motion forecasting baselines,
(2) accident anticipation/detection datasets, and
(3) V2X-enabled cooperative benchmarks.
It highlights each dataset’s scenario (ego, infrastructure, or multi-agent),
source (real-world or simulator),
available sensing modalities (e.g., LiDAR, radar, gaze, vehicle states, semantics),
and supported benchmark tasks such as detection, tracking, motion forecasting, and accident anticipation.
\\
\hline
\end{tabular}
\end{table*}

Robust cooperative perception relies on the ability to fuse heterogeneous information from multiple sensors and agents into a coherent and reliable scene representation. 
In safety-critical driving, no single modality—camera, radar, LiDAR, or V2X—can provide complete situational awareness under various conditions. 
\emph{Multi-modal fusion} integrates complementary cues across sensing types, while \emph{multi-agent fusion} integrates these cues across viewpoints, collectively forming a perception field that is more complete, resilient, and temporally consistent than any individual input~\cite{arnold2020cooperative, morais2025deepsense}. 
Such fusion underpins both short-term motion forecasting and early accident anticipation, providing redundancy against occlusions, sensor noise, and communication dropouts.

From the system perspective, the perception layer fuses not only features but also their contextual attributes—source identity, time-stamp, and reliability tags—thus constructing a probabilistic semantic world model ready for risk reasoning. 
Recent frameworks couple this fusion with communication state awareness, using transformer-based encoders that attend jointly to visual, radar, and V2X features while adapting the data rate and message priority to link quality~\cite{gharsallah2024mvx,liu2019edge}. 
By embedding both physical (network) and semantic (scene) conditions into a single reasoning loop, these systems maintain calibrated uncertainty, allowing downstream modules to distinguish genuine threats from spurious detections and to set intervention thresholds accordingly~\cite{zha2025heterogeneous, sheng2025semantic}. 
In essence, fusion transforms fragmented, asynchronous observations into structured, forecast-ready evidence.

The recent growth of cooperative datasets has made this fusion paradigm empirically verifiable. 
Table~\ref{tab:dataset_comparison} summarizes the progression from ego-centric datasets to fully cooperative, multi-modal benchmarks. 
\textbf{OPV2V}~\cite{xu2022opv2v} introduced simulated LiDAR-based V2V perception with feature- and object-level sharing under dynamic topology, 
demonstrating how multi-agent fusion improves 3D detection beyond single-vehicle baselines. 
\textbf{DAIR-V2X}~\cite{yu2022dair} and \textbf{V2X-Seq}~\cite{yu2023v2x} extended this idea to real-world V2I scenarios, aligning camera–LiDAR–RSU views across space and time to benchmark synchronization and calibration quality. 
\textbf{V2X-Sim}~\cite{li2022v2x} and \textbf{V2X-Seq/Forecasting} further connected perception to prediction, providing multi-agent sequences suitable for cooperative motion forecasting and cross-agent temporal alignment. 
More recently, \textbf{DeepSense-V2V}~\cite{morais2025deepsense} incorporated not only visual and LiDAR signals but also communication-state metadata—latency, packet loss, and signal strength—linking perception fidelity directly to network dynamics.

Despite this progress, notable gaps remain. 
Few datasets explicitly annotate PQoS indicators or provide joint labels for perception, forecasting, and risk inference, 
and synchronization under realistic bandwidth and latency constraints is still underrepresented~\cite{coll2022end, zhang2024latency}. 
Future benchmarks may therefore integrate sensor diversity, agent heterogeneity, and communication metrics to form end-to-end safety datasets that capture both perceptual evidence and its temporal validity.

Together, these benchmarks provide the empirical substrate for multi-modal and multi-agent fusion: 
they enable reproducible evaluation, cross-domain calibration, and systematic comparison across cooperation levels. 
By linking multi-sensor fusion to communication-aware design, these datasets ground perception as the bridge between raw evidence and forecast-ready understanding, ensuring that cooperative intelligence in V2X systems is both measurable and actionable.

\subsubsection{Timeliness, Synchronization, and PQoS Alignment}

Safety-critical perception delivers value when cooperative evidence arrives in time to shape decisions. 
At highway speeds, a delay of tens of milliseconds can determine whether a predicted maneuver remains actionable. 
Therefore, timeliness and synchronization form the temporal backbone of the SPD loop: 
they ensure that every observation, fusion, and prediction refers to the same physical moment and is received before its relevance expires~\cite{gyawali2020challenges,zhou2020evolutionary}. 

Empirical evaluations confirm this sensitivity. 
On the DAIR-V2X-C benchmark, perception advantages degrade sharply once infrastructure-to-vehicle latency exceeds $100$\,ms, 
effectively reverting cooperative models to ego performance~\cite{yu2022dair,zhang2025iot}. 
Similar studies show that cooperative inference benefits plateau when perception-to-decision delay exceeds the motion horizon of fast-moving targets~\cite{coll2022end,zhang2024latency}. 
Such evidence establishes latency not merely as a networking metric but as a determinant of safety utility and trustworthiness.

Modern V2X architectures mitigate these temporal challenges through edge-assisted scheduling and \emph{PQoS} regulation. 
MEC at RSUs or micro data centers brings perception and decision services closer to vehicles, 
eliminating long backhaul loops and sustaining sub-$20$\,ms application-level round trips~\cite{coll2022end}. 
By hosting cooperative fusion and inference directly at the edge, MEC enables near-synchronous updates of shared features and risk maps, 
while PQoS metadata---including latency, freshness, and loss indicators---allow downstream forecasting modules to calibrate how much to trust each input~\cite{hu2025trust,hu2025toward}. 
This design converts network variability into a quantified, contract-like property: 
each packet of cooperative evidence carries its temporal validity and confidence envelope, ensuring that decision modules act on recent, credible information.

Synchronization complements timeliness by aligning observations across agents and communication tiers. 
NR-V2X and 5G-based sidelink frameworks provide sub-millisecond clock coherence through GNSS-assisted and network-assisted synchronization~\cite{saad2021advancements,staubach2009factors}. 
Hardware-in-the-loop validation further shows that edge schedulers can preserve this alignment even under congested conditions~\cite{zhang2024latency}. 
Such temporal coherence is important for fusing trajectories, occupancy grids, and risk estimates that originate from spatially distributed sensors yet need to refer to the same physical instant.

Ultimately, the integration of timeliness control, synchronization, and PQoS alignment forms the temporal contract of cooperative perception. 
Within the SPD framework, these mechanisms translate heterogeneous and asynchronous streams into temporally consistent, deadline-aware evidence that forecasting and intervention modules can consume. 
By embedding timing awareness into both communication and computation, cooperative safety intelligence evolves from best-effort connectivity to time-bounded reasoning---a prerequisite for proactive and verifiable accident prevention~\cite{golias2002classification,hu2025toward}.

\subsubsection{Safety-Ready Outputs, PQoS, and the Evidence Contract}

{Cooperative perception becomes safety-relevant when shared geometric and semantic outputs are transformed into decision-consumable evidence.} 
Safety-critical perception therefore culminates in \emph{forecast-ready artifacts}---structured, time-stamped, and uncertainty-aware representations that downstream forecasting and risk modules can directly consume~\cite{chang2023bev,wang2024bevgpt}. 
Within cooperative systems, these artifacts constitute a contractual interface between perception and decision: each output is accompanied, where available, by descriptors of spatial alignment, temporal validity, and confidence calibration~\cite{hu2025trust,peng2021evaluation}.
{By organizing raw detections, BEV maps, trajectories, and occupancy estimates into a evidence format, the perception layer enables safety reasoning over interpretable, verifiable, and temporally valid inputs~\cite{chang2023bev,yu2023v2x}.}

{Within this evidence contract, PQoS denotes the operational quality of perception-derived safety evidence after sensing, communication, and fusion have jointly acted on it~\cite{ren2024interruption,lin2024edge}. It differs from conventional link-level QoS by describing whether an evidence item is sufficiently fresh, reliable, calibrated, and valid for downstream decision-making~\cite{mason2025prata,stenhammar2025clustering}. In this survey, PQoS is characterized through latency, freshness, packet or feature loss, source provenance, uncertainty calibration, and validity window~\cite{morais2025deepsense,cai2024cooperative}. These descriptors support trust calibration, weighting, timing, and suppression decisions for received objects, trajectories, occupancy maps, and risk cues.}

\begin{figure*}[t]
    \centering
    \includegraphics[width=0.9\linewidth]{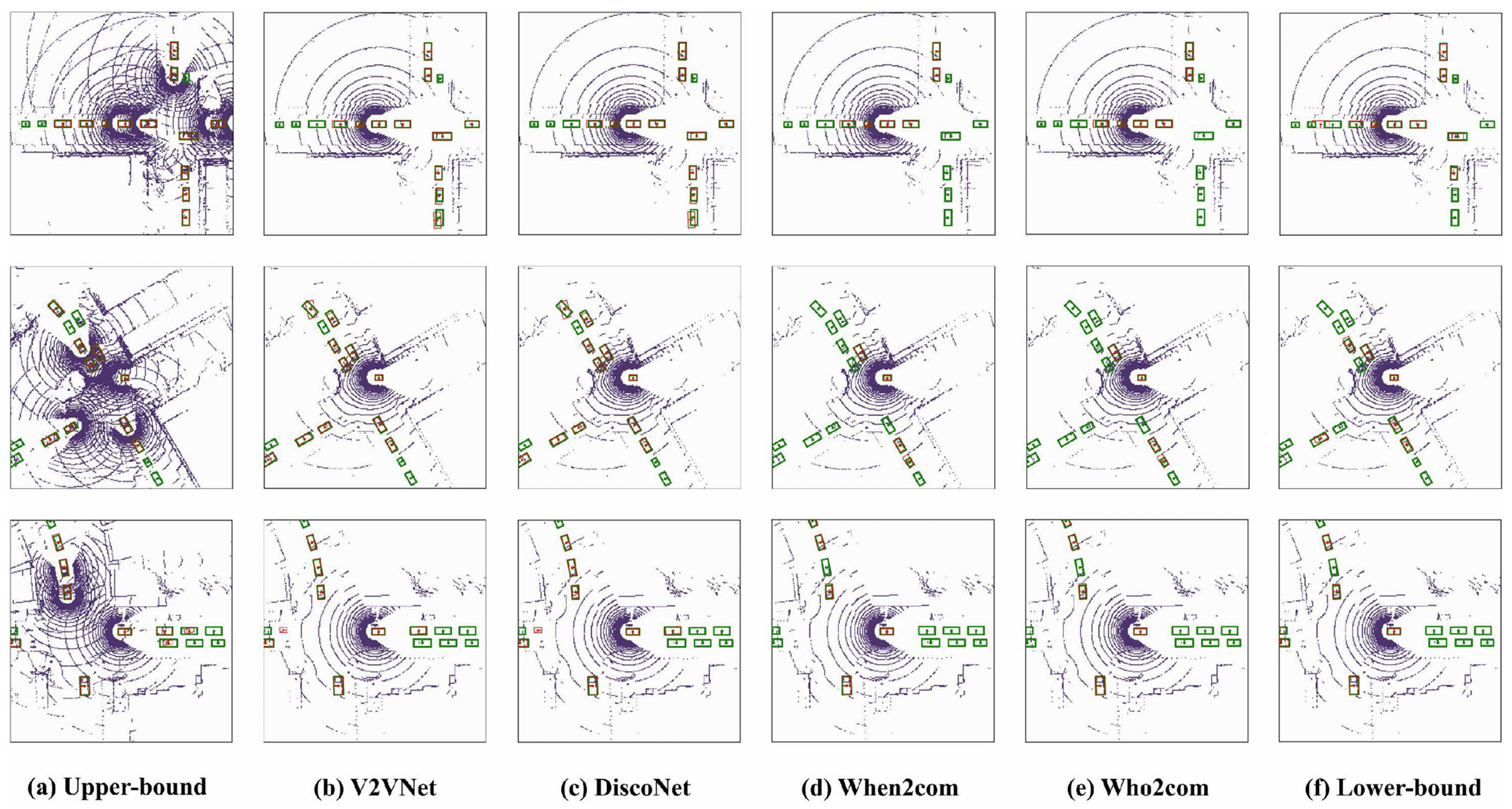}
    \caption{Example of BEV-based cooperative perception in \textit{V2X-Sim}. 
    Gray points denote LiDAR data from RSUs, while colored clusters correspond to vehicle-mounted sensors. 
    Orange boxes mark 3D bounding boxes of detected agents~\cite{li2022v2x}. 
    The BEV fusion unifies multi-agent, multi-view observations into a structured, temporally aligned representation ready for forecasting.}
    \label{fig:bev}
    \vspace{-5mm}
\end{figure*}

\textbf{Structure, Timestamps, and Trust:}
This evidence contract is maintained through message and format standards such as BSM, SPaT, and RSM, which translate cooperative perception into forms consumable by planners and safety evaluators. 
Each message bundle integrates three essential properties: 
(1)~\textbf{Structure}—a unified spatial frame that aligns multi-agent observations into consistent coordinates; 
(2)~\textbf{Timestamping}—synchronized clocks and PQoS tags that quantify latency, freshness, and sampling uncertainty; and 
(3)~\textbf{Trust metadata}—cryptographic credentials, calibration quality, and privacy-preserving mechanisms ensuring data integrity and accountability~\cite{hasan2020securing,huang2020recent,yoshizawa2023survey,bute2023trust,sharma2022security}. 
Together, these elements convert heterogeneous sensor outputs into operationally traceable evidence, elevating cooperative perception from best effort to verifiable reliability.

\textbf{Forecasting as Structured Evidence:}
For proactive safety, forecasts can be organized as \emph{structured evidence bundles}—probabilistic occupancies, multi-hypothesis flows, and interaction graphs annotated with timestamps and PQoS metadata.
This framing enables prediction models to quantify both what is likely to happen and how long the underlying evidence remains valid. 
Benchmark datasets such as \textbf{V2X-Sim}, \textbf{V2X-Seq}, and \textbf{DeepAccident} illustrate this transition: 
BEV fusion integrates RSU and vehicle perspectives into spatially coherent grids for cooperative forecasting~\cite{li2022v2x,chang2023bev}, 
while DeepAccident expresses predictions as calibrated collision likelihoods linked to explicit time and location outcomes~\cite{almutairi2025deep}. 
These cases exemplify the evidence contract in operation, where forecasts become risk-aware statements grounded in temporally consistent and uncertainty-quantified perception.
Within the SPD framework, the perception layer transforms raw and asynchronous observations into temporally aligned, trust-annotated evidence, turning cooperative data into actionable foresight for downstream decisions.

\subsection{Decision Layer for Safety: From Calibrated Risk to Coordinated Action}

The decision layer represents the culmination of the SPD loop, where calibrated perception evidence is transformed into policies and actions that safeguard cooperative traffic. Building on shared perception, this layer determines \emph{how and when to act} based on temporally aligned, uncertainty-aware, and context-tagged evidence. Through risk inference, multi-agent consensus, and adaptive intervention, the decision process helps distributed awareness evolve into coordinated behavior across connected agents. By embedding PQoS constraints, human interpretability, and fallback continuity, the decision layer operationalizes the transition from shared foresight to executable safety intelligence, turning knowledge into timely, consistent, and trustable action.

\subsubsection{From Calibrated Evidence to Actionable Risk}

\begin{figure}[t]
    \centering
    \includegraphics[width=0.4\textwidth]{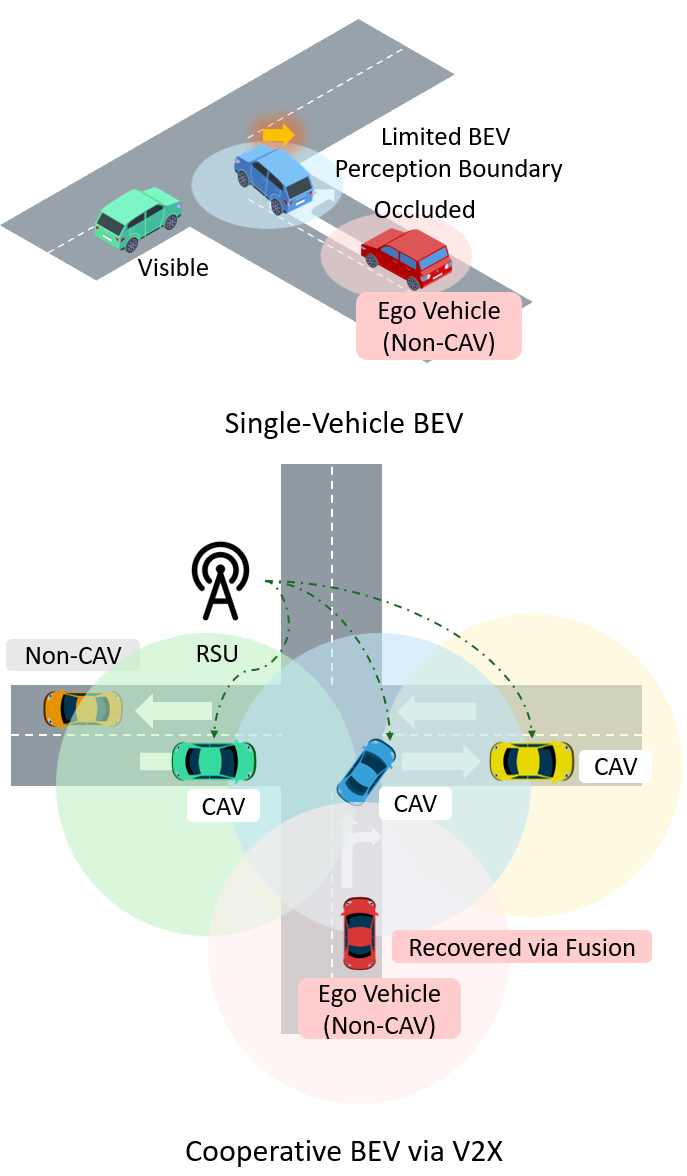}
    \caption{Example of cooperative perception outputs expressed in BEV-V2X form. The ego-only BEV (left) shows limited coverage, while the cooperative BEV (right) fuses multi-vehicle and RSU inputs into a unified representation that forecasting models can directly consume as structured evidence~\cite{chang2023bev}.}
    \label{fig:bev_coop}
    \vspace{-5mm}
\end{figure}

The decision layer interprets calibrated, time-stamped evidence from perception to form an \emph{actionable understanding of risk}. Rather than treating detection or forecasting as terminal outputs, SPD views them as structured inputs for reasoning. Each forecast-ready bundle—comprising multimodal trajectories, occupancy probabilities, uncertainty calibration, and PQoS descriptors—serves as a semantic contract between perception and decision. Building upon this evidence, risk inference proceeds through a continuous reasoning pipeline that integrates geometric feasibility, probabilistic calibration, and cost-aware timing into a unified decision process.

First, \emph{geometric and topological feasibility} establishes whether agents’ safety envelopes will intersect under maneuvers in the near future. Intersection phase and timing information (SPaT/MAP), right-of-way rules, and BLOS cues from RSUs define when and where such violations become plausible \cite{wegener2021longitudinal,liu2024eco}. These checks produce deterministic constraints that frame the subsequent probabilistic reasoning. Second, \emph{probabilistic inference} integrates calibrated forecasts with contextual priors derived from source quality—coverage, consistency, and historical calibration performance—to produce comparable likelihoods across viewpoints \cite{li2024spatiotemporal,shi2024mtr,chang2023bev}. Cooperation reduces epistemic uncertainty by merging partial evidence from ego and infrastructure perspectives, ensuring that a ``0.3 collision likelihood'' carries the same operational meaning system-wide. Third, \emph{cost- and timing-aware judgment} transforms likelihood into decision readiness by evaluating the expected safety gain, human response latency, and secondary risk of intervention in dense traffic. Shared ``cost landscapes,'' disseminated via V2X links, align braking or lane-change thresholds among nearby agents \cite{ribeiro2023evaluation,kubra2024integrated}. 

\begin{figure}
    \centering
    \includegraphics[width=1\linewidth]{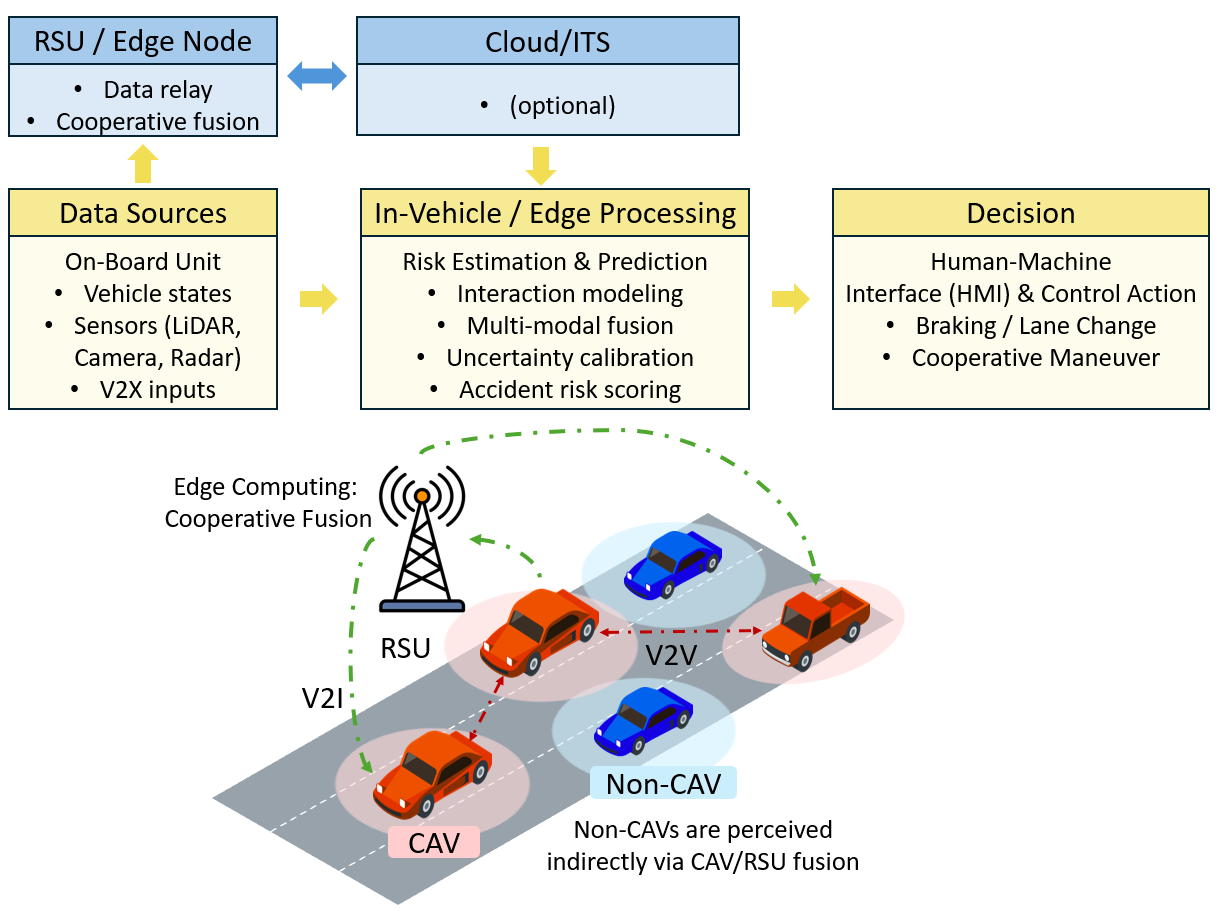}
    \caption{Overall V2X-enabled crash risk prediction framework. Vehicle interaction data are collected onboard, processed to estimate and predict crash risk, and shared via V2V/V2I links to trigger coordinated warnings \cite{jo2022vehicle}.}
    \label{fig:overall_v2x_risk}
    \vspace{-5mm}
\end{figure}

Fig.~\ref{fig:bev_coop} and Fig.~\ref{fig:overall_v2x_risk} illustrate this inference process in both spatial and functional views. The cooperative BEV representation (Fig.~\ref{fig:bev_coop}) fuses multi-vehicle and RSU inputs to recover occluded agents, while the end-to-end risk prediction framework (Fig.~\ref{fig:overall_v2x_risk}) depicts how calibrated evidence flows from distributed sensing to coordinated warning generation. Together, they show how SPD transforms what is merely \emph{visible and knowable} into what becomes \emph{decidable and executable}, producing decision-level risk tokens that couple foresight with timing and cost awareness.

\subsubsection{Coordinated Risk Aggregation and Triggering under PQoS Constraints}

Within the SPD loop, the decision layer consolidates asynchronous and sometimes conflicting risk estimates into a coherent, trust-calibrated state to support subsequent intervention. This consolidation connects \emph{distributed risk aggregation} with \emph{PQoS-aligned triggering}, helping align cooperative actions with evidence timing and quality. Each agent contributes locally calibrated forecasts enriched with uncertainty, timestamps, and PQoS metadata, which are aggregated through soft weighting and consensus rules to keep the system responsive and stable under partial cooperation.

Source-weighted fusion assigns dynamic trust scores to each contributing viewpoint according to its coverage, historical calibration reliability, and communication quality. Degraded streams receive lower weights, preserving minority but valuable observations such as an RSU's view of an occluded pedestrian~\cite{liu2024eco,chang2024cav}. This soft aggregation builds consensus over time as multiple independent perspectives corroborate a shared event hypothesis. When agreement emerges, the decision layer synthesizes a compact and consistent risk representation---expressed as a token containing actor identity, location, urgency, and temporal validity---ready to drive coordinated triggers.

\begin{figure*}[t]
    \centering
    \includegraphics[width=0.9\textwidth]{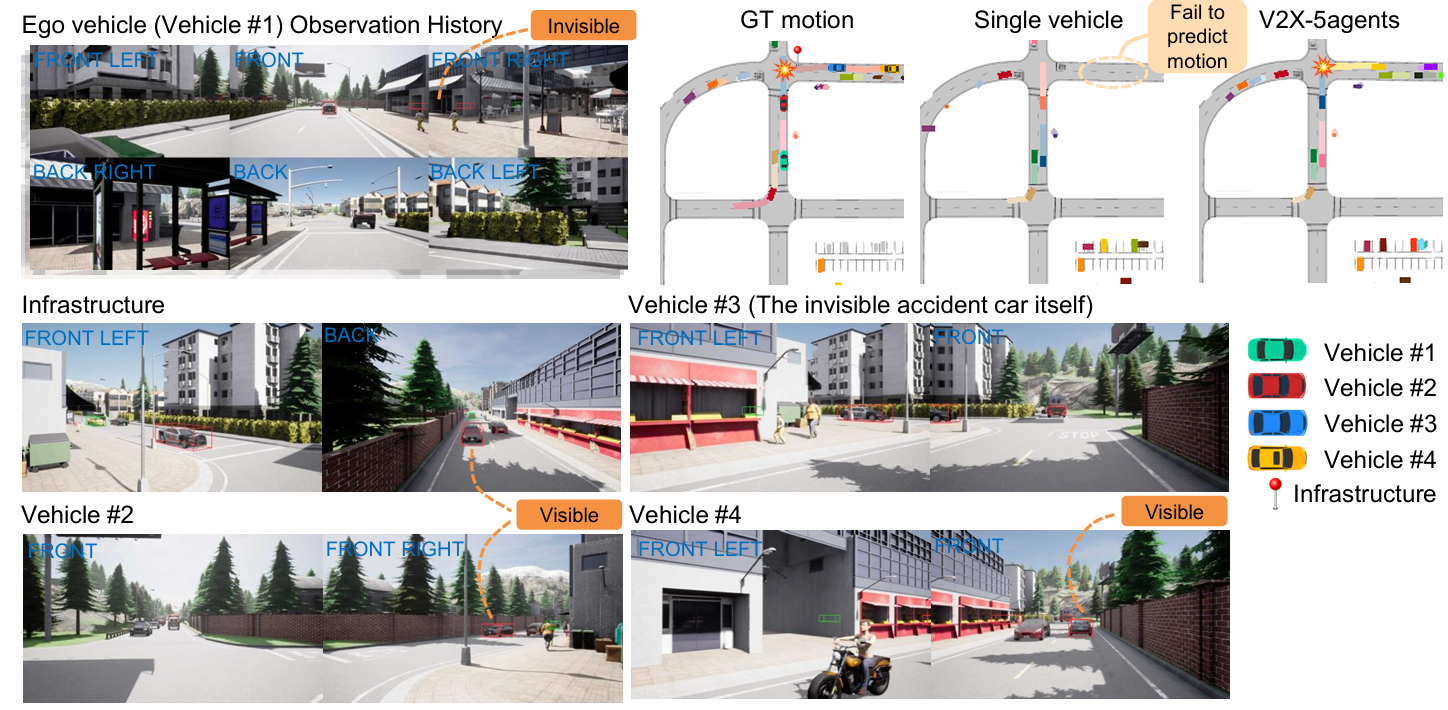}
    \caption{Illustration of the DeepAccident end-to-end motion and accident prediction task \cite{wang2024deepaccident}. The ego vehicle (Vehicle~\#1) relies solely on its own cameras and cannot perceive the hidden Vehicle~\#3 due to building occlusion, leading the single-vehicle model (center, ``Single vehicle'') to miss the impending collision. In contrast, the V2X-5agents model fuses views from multiple cooperating vehicles and infrastructure, reconstructing the motion of the occluded Vehicle~\#3 and correctly predicting the accident. This demonstrates how cooperative viewpoints transform an ``unseen'' hazard into an actionable risk.}
    \label{fig:deepaccident_case}
    \vspace{-5mm}
\end{figure*}

PQoS acts as a gating and scheduling mechanism for these triggers. Latency and evidence degradation shape the actionability of alerts; stale or incomplete evidence is assigned advisory intensity to support smoother control escalation~\cite{coll2022end}. The \emph{commitment window}, defined by reaction time and braking envelopes, establishes the time range in which a trigger can influence the outcome. Within this window, cooperative evidence extending beyond line of sight lengthens anticipation and improves recall of rare events, as demonstrated in accident-oriented pipelines such as \emph{DeepAccident} (Fig.~\ref{fig:deepaccident_case})~\cite{wang2024deepaccident,almutairi2025deep}. Together, PQoS gating and commitment-window timing transform probabilistic predictions into temporally disciplined actions, supporting early intervention as a designed property of cooperative safety intelligence.

Cooperation further reduces redundant or conflicting interventions through lightweight consensus. Vehicles and RSUs exchange risk tokens to coordinate action, yielding, and confirmation roles. RSUs, when present, can arbitrate intersection-level conflicts and broadcast unified instructions; in pure V2V settings, deterministic leader selection---based on distance to hazard, visibility, or trust weight---helps propagate triggers~\cite{jo2022vehicle,bejarbaneh2024exploring}. 
When communication quality degrades, systems can operate with local geometric checks and conservative defaults, maintaining safety continuity under partial connectivity~\cite{li2023learning,aung2023deep}.
Fig.~\ref{fig:bev_coop} visualizes how cooperative perception expands the decision layer's evidence horizon, while Fig.~\ref{fig:overall_v2x_risk} situates these mechanisms within the complete sensing-to-decision loop. Together, they illustrate how SPD aligns probabilistic calibration, temporal coherence, and coordination logic under PQoS constraints, turning distributed foresight into synchronized, human-aware action.

\subsubsection{Human-Centered Intervention and System Recovery under SPD Integrity}

In cooperative safety intelligence, the human-in-the-loop remains the ultimate interpreter and executor of SPD decisions. The decision layer is expected to ensure that every alert or intervention is not only algorithmically valid but also \emph{usable, interpretable, and trust-calibrated} by human operators. Usability and robustness form a dual imperative: the former guarantees that drivers or supervisors can act correctly within the available time window, while the latter ensures that cooperative functions degrade and recover smoothly when network or sensor quality fluctuates. Together, they transform SPD from a closed technical loop into an auditable human–machine–environment feedback cycle.

\begin{figure*}[t]
    \centering
    \includegraphics[width=0.8\linewidth]{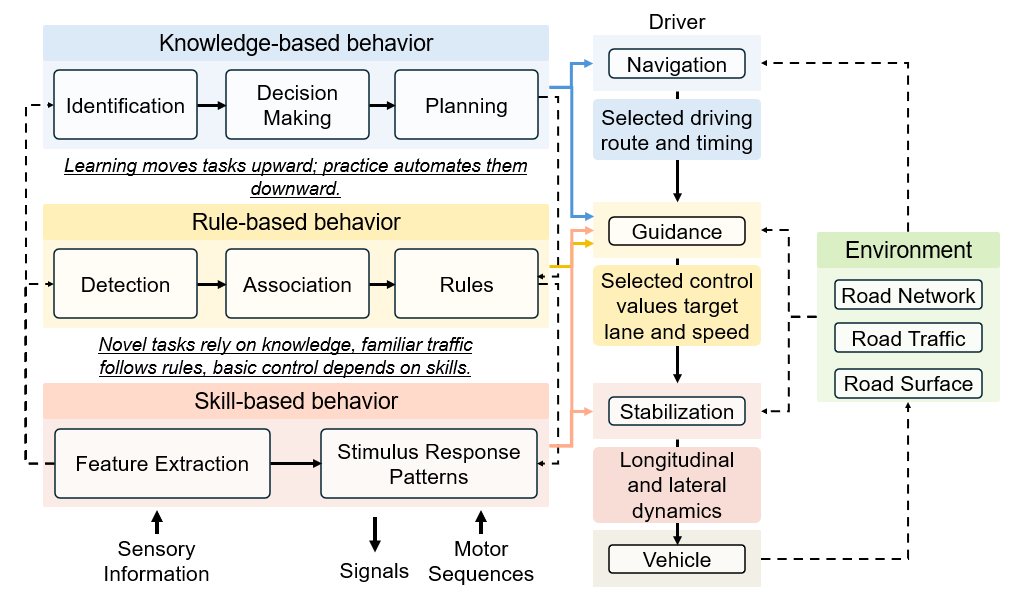}
    \caption{Human factors in driving tasks: simplified Wiener’s model linking knowledge-, rule-, and skill-based behaviors to vehicle control~\cite{olaverri2017human}. The model illustrates how information flows from sensory inputs through cognitive layers, producing motor sequences that guide vehicle dynamics. Novel tasks rely on knowledge, familiar traffic follows rules, and basic control depends on automated skills.}
    \label{fig:wiener_model}
    \vspace{-5mm}
\end{figure*}

\textbf{Graded human-centered alerts.}  
Effective cooperative warnings often take a staged form: \emph{inform} (contextual cue), \emph{warn} (focused highlight), and \emph{command} (explicit instruction). A typical message includes the \emph{object}, \emph{direction}, and \emph{time-to-arrival (TTA)}, complemented by confidence and age of the underlying evidence. According to Wiener’s hierarchy (Fig.~\ref{fig:wiener_model}), early and clearly localized cues allow knowledge- and rule-based reasoning (e.g., yielding or route re-planning), while late-stage triggers compress into skill-based reflexes. Within the SPD loop, this framing highlights the value of earlier cues for higher-level cognitive engagement, while still keeping alerts usable when latency or workload rises.

\textbf{Trust and transparency in cooperative cues.}  
Human acceptance depends on both \emph{source transparency} and \emph{temporal credibility}. PQoS metadata—such as timestamp, latency, and packet integrity—serve as the bridge between machine reliability and human trust. Alerts that reveal their evidence age and confidence invite appropriate skepticism without confusion. RSU-mediated consensus ensures that cooperating agents deliver consistent and non-contradictory warnings, which directly supports trust calibration and avoids ``swarm braking.'' Multimodal presentation (visual, auditory, haptic) aligns with situational urgency and driver attention limits, translating SPD outputs into cognitively compatible signals~\cite{kubra2024integrated,jo2022vehicle}. 

\textbf{Graceful degradation and recovery.}  
When connectivity degrades, the system transitions along a controlled fallback ladder: (1) reduce feature fidelity and compress transmissions; (2) switch to object- or occupancy-level tokens; (3) revert to ego-centric geometric checks with enlarged safety margins; and (4) upon reconnection, rejoin the cooperative state with hysteresis to prevent oscillation~\cite{li2024spatiotemporal,zhang2024latency}. MEC-assisted edge nodes near RSUs shorten restoration times and preserve temporal alignment. Each transition is logged through PQoS traces, allowing verification of degradation duration, suppression ratio, and recovery latency, thereby making robustness measurable through explicit operational records.

\begin{table*}[t]
\caption{Safety-critical application requirements~\cite{clancy2024wireless}, with human-centered considerations under SPD.}
\label{tab:safety_requirements}
\centering
\small
\setlength{\tabcolsep}{6pt}
\renewcommand{\arraystretch}{1.3}
\rowcolors{2}{yellow!10}{yellow!3}
\resizebox{0.95\textwidth}{!}{%
\begin{tabular}{l c c c m{4.8cm}}
\rowcolor{yellow!25}
\hline
\textbf{Use Case} & \textbf{V2X Type} & \textbf{Max. Latency} & \textbf{Min. Freq.} & \textbf{Human-centered Considerations} \\
\hline

\rowcolor{gray!15}
\multicolumn{5}{l}{\textit{Cooperative Awareness}} \\

Emergency vehicle warning         & V2V/I   & 100\,ms & 10\,Hz & Direction, urgency, early yielding \\
Slow vehicle indication           & V2V/I   & 100\,ms & 2\,Hz  & Advisory, smooth deceleration \\
Pre-crash sensing warning         & V2V/I   & 20\,ms  & 10\,Hz & Reflex-level, low-jerk braking \\
Forward collision warning         & V2V     & 100\,ms & 2\,Hz  & Object-specific, TTA clarity \\
Intersection collision warning    & V2V/I   & 100\,ms & 10\,Hz & SPaT/MAP context, right-of-way rationale \\
VRU warning (pedestrian/cyclist)  & V2V/I/P & 100\,ms & 1\,Hz  & Conservative cues, direction + distance \\
Lane change assistance            & V2V/I   & 100\,ms & 10\,Hz & Rule-based guidance, blind-spot clarity \\
Overtaking vehicle warning        & V2V/I   & 100\,ms & 10\,Hz & Deduplication, single-source trigger \\

\rowcolor{gray!15}
\multicolumn{5}{l}{\textit{Road Hazard Warnings}} \\

Emergency brake lights            & V2V/I   & 100\,ms & 10\,Hz & Graded braking, comfort preservation \\
Wrong-way driving warning         & V2V/I   & 100\,ms & 10\,Hz & Salience, approach direction, timing \\
Roadwork warning                  & V2V/I   & 100\,ms & 2\,Hz  & Zone geometry, merge stability \\
Collision risk warning            & V2V/I   & 100\,ms & 10\,Hz & Object+direction, trust calibration \\
\hline
\vspace{-5mm}
\end{tabular}}
\end{table*}

\textbf{Integrating fallback and human response.}  
Fallback is not merely technical—it also has perceptual and behavioral consequences. During degradation, the system communicates its operational mode and confidence level so that the driver’s situational model stays synchronized. When the cooperative state is restored, recovery messages include a short hysteresis window to prevent overreaction or oscillation. Human and machine thus share a mutual predictability: the human understands what level of assistance is currently active, and the machine anticipates how the driver will respond under workload or stress. Such mutual transparency, supported by PQoS auditing, is central to SPD’s aim of building trustworthy, self-explaining safety intelligence.

\subsection{Lessons Learned from Communication-Aware Cooperative Perception}
{From the above discussion, cooperative perception in V2X-enabled safety systems provides its main value by complementing local observations and generating more consistent scene evidence under occlusion, BLOS, and multi-view uncertainty. This value is shaped by how information is exchanged among agents, since early, intermediate, and late fusion reflect different balances among information richness, communication cost, synchronization demand, and correction capability. In practical deployment, communication efficiency, latency tolerance, pose alignment, and heterogeneous sensing need to be considered together with perception accuracy, as these factors determine whether cooperative outputs remain timely and reliable. Future evaluation of cooperative perception would therefore benefit from a system-level perspective that jointly reports geometric accuracy, communication cost, latency behavior, calibration quality, robustness under degraded links, and deployment conditions.}

\section{SPD-Aligned Datasets, Benchmarks, and Platforms}
\label{sec:spd_datasets_bench_platforms}

This section frames datasets, metrics, and platforms around cooperation \emph{granularity} and \emph{closed-loop} evaluation to make SPD claims measurable and comparable. We classify datasets by evidence scope, define evaluation criteria beyond geometry—covering \emph{earliness}, \emph{calibration}, and \emph{fleet-level coordination}—and examine co-simulation and digital-twin platforms for realistic end-to-end testing.

\subsection{Dataset Landscape by Cooperation Granularity}
\label{subsec:dataset_landscape}

Datasets aligned with SPD differ not only by sensors and labels but also by the \emph{cooperation granularity} they support—what viewpoints are synchronized, how provenance and PQoS are recorded, and whether sequential context enables forecasting-to-decision studies.

\textbf{Single-vehicle baselines.}
Canonical datasets such as KITTI, nuScenes, and Waymo provide indispensable annotations for detection, tracking, and motion forecasting. They form the foundational benchmarks for evaluating perception and trajectory prediction quality, yet their scope remains limited to ego-centric viewpoints, leaving occluded or BLOS scenarios and multi-view consistency largely unexplored.

\textbf{V2V/V2I cooperative datasets.}
V2X\textnormal{-}Sim~\cite{li2022v2x} (CARLA+SUMO) exposes \emph{controlled} multi-vehicle/RSU streams (cameras/LiDAR) with tight time/pose alignment and configurable link conditions, making it suitable for isolating how bandwidth, loss, and partner selection shape \emph{forecast-ready} evidence (Fig.~\ref{fig:bev}). 
V2X\textnormal{-}Seq~\cite{yu2023v2x} extends to \emph{real-world} sequential RSU–vehicle scenes (hours-long, intersection-rich), enabling studies of occlusion recovery, calibration drift, and redundancy suppression over time.

\textbf{End-to-end safety pipelines.}
DeepAccident~\cite{wang2024deepaccident} illustrates a full chain from cooperative perception to accident anticipation via multi-view fusion and risk scoring; it is particularly informative for BLOS cases where ego models miss hidden actors (Fig.~\ref{fig:deepaccident_case}). 
Across these resources, SPD gaps persist in standardized PQoS/provenance fields, accident and intervention tags aligned with forecasts, and synchronized multi-agent labels for auditing \emph{who saw what, when}.

\subsection{Evaluation: Beyond Geometry (Earliness, Calibration, Coordination)}
\label{subsec:evaluation_beyond_geometry}

Classical geometric metrics (e.g., mAP, displacement errors) are necessary but insufficient for the SPD interface, where a predictor is valuable insofar as it produces \emph{actionable, calibrated, and consistent} risk under realistic communication. We adopt a three-part contract.

\textbf{Earliness.}
Mean Time-to-Arrival of the first correct alert (mTTA) and TTA@R quantify decision headroom at fixed recall, discouraging myopic ``early-but-wrong'' operating points. 
Early-utility curves further visualize the trade-off between earliness and false-alarm cost for deployment tuning. Accident-aware utility (e.g., APA) is particularly informative when reported together with mTTA~\cite{wang2024deepaccident}.

\textbf{Calibration.}
ECE/ACE, Brier score, and NLL evaluate probability quality so that the same threshold has the same meaning across sources. 
Since evidence quality depends on links, calibration numbers are \emph{conditioned} on communication regimes (latency/loss/compression) used during data collection or simulation~\cite{coll2022end}. 
This conditioning turns raw accuracy into \emph{auditable} readiness for triggering.

\textbf{Coordination and rare events.}
Fleet consistency is measured via inter-agent agreement, oscillation/suppression rates, and a ``no-swarm-braking'' score that penalizes redundant triggers~\cite{jo2022vehicle,bejarbaneh2024exploring}. 
Given severe class imbalance, minority-focused metrics are reported explicitly, including PR-AUC for rare buckets and recall at fixed low-FPR operating points~\cite{richardson2024receiver}.
All results are accompanied by timestamps, source identifiers, and PQoS tags to maintain end-to-end traceability.

\subsection{Co-Simulation \& Digital Twins for Closed-Loop Testing}
\label{subsec:cosim_digital_twins}

Closed-loop assessment requires platforms that couple sensing, perception, communication, and decision with reproducible perturbations.

\textbf{Communication-aware co-sim.}
V2X\textnormal{-}Sim~\cite{li2022v2x} (and follow-ups~\cite{liu2025towards}) enables controlled variation of bandwidth budgets, packet loss, message rates, and pose noise. 
Such knobs let researchers trace how compression or drops propagate into \emph{calibration drift} and \emph{time-to-alert} changes, and whether de-duplication/consensus policies prevent ``alert storms'' under partial views (Fig.~\ref{fig:bev}).

\textbf{Digital twins with sequential realism.}
V2X\textnormal{-}Seq~\cite{yu2023v2x} provides long-horizon, intersection-rich RSU–vehicle sequences where occlusions, phase timing, and traffic variability naturally create BLOS and confirmation delays. 
This supports auditing of recovery after outages, calibration over drift, and stability of repeated interventions.

\textbf{End-to-end safety case studies.}
Pipelines like DeepAccident~\cite{wang2024deepaccident} close the loop from multi-view BEV fusion to accident anticipation, highlighting how cooperative viewpoints extend anticipation horizons and raise alert credibility in dense urban scenes (Fig.~\ref{fig:deepaccident_case}). 
For SPD alignment, recording PQoS traces, synchronization status, and source provenance supports ex-post verification of trigger decisions.

\subsection{Lessons Learned from SPD-Based Cooperative Safety Intelligence}
{From the SPD perspective, cooperative safety intelligence can be viewed as a gradual transformation from sensing evidence to perception evidence and then to decision-support evidence. This process highlights the importance of making safety-ready outputs both informative and usable: in addition to object, trajectory, or occupancy estimates, such outputs can carry timing, uncertainty, provenance, calibration, and PQoS-related information whenever available. These attributes help downstream risk reasoning interpret urgency, validity windows, and possible intervention consequences. At the decision level, coordinated triggering, fallback, and recovery mechanisms provide practical support when connectivity, evidence quality, or agent agreement becomes uncertain. Human-centered intervention further emphasizes the compatibility of safety intelligence with driver response time, cognitive workload, and trust calibration. Taken together, these lessons point to SPD-based V2X safety systems that jointly consider evidence quality, timing, confidence, coordination, and human usability across the cooperative safety loop.}
 
\section{Future Directions for the SPD Framework}

The proposed SPD framework provides a unifying paradigm for cooperative safety intelligence, yet its current form remains primarily conceptual. Moving SPD from an organizing principle toward a deployable system architecture calls for sustained investigation across several directions. These directions span scalable data infrastructures, embodied predictive reasoning, and trustworthy human-in-the-loop cooperation, and connect algorithmic progress with real-world deployment and societal validation.

\subsection{Scalable SPD Data and Evaluation Infrastructure}
\label{subsec:spd_data_eval}

Achieving \emph{safety-ready} cooperation along the SPD loop requires data and evaluation infrastructure that go beyond ego geometry and per-frame labels. SPD needs (i) synchronized, sequential, multi-agent recordings that preserve cross-view \emph{provenance} and timing, (ii) benchmarks that expose how perception evidence degrades or recovers under realistic networking conditions (latency, loss, pose drift), and (iii) \emph{contract-style} metadata that travels with each artifact from Sensing to Perception to Decision.

\subsubsection{Sequential, multi-agent, and infrastructure-supported sensing}
V2X\textnormal{-}Seq establishes a large-scale, sequential V2X corpus with vehicle and infrastructure streams, vector maps, and traffic-light signals; it introduces VIC3D tracking and both online/offline VIC forecasting tasks, moving from framewise perception toward temporally extended forecasting under cooperation~\cite{yu2023v2x}. It highlights the gap left by single-vehicle benchmarks and positions itself as a large-scale sequential V2X dataset captured from natural scenes~\cite{yu2023v2x}. Meanwhile, canonical single-vehicle sets (KITTI, nuScenes, Waymo, etc.) remain indispensable but inherently ego-centric; V2X\textnormal{-}Seq explicitly contrasts with infrastructure-only or simulation-only corpora to motivate multi-view, multi-agent evidence~\cite{yu2023v2x}.

\subsubsection{Communication-aware, controllable digital twins}
On the synthetic side, V2X\textnormal{-}Sim couples SUMO micro-traffic with CARLA to generate tightly synchronized multi-vehicle and RSU streams, and provides an open-source testbed and benchmarks for collaborative detection, tracking, and segmentation—precisely the knobs needed to stress cooperation logic repeatably~\cite{li2022v2x}. By design, such platforms make the communication substrate \emph{first-class}: bandwidth, message rate, and sensor-view composition can be varied while holding scene semantics constant, enabling fair audits of perception robustness under controlled link conditions~\cite{li2022v2x}.

\subsubsection{Provenance, synchronization, and PQoS-style metadata}
DeepSense-V2V shows how to co-collect multi-modal sensing and mmWave communication with panoramic coverage and programmatically generated labels, while tracking sequence continuity and alignment across modalities~\cite{morais2025deepsense}. Beyond content labels, it attaches timing, index, and link descriptors—illustrating the \emph{evidence metadata} required by SPD for assessing freshness, trust, and completeness~\cite{morais2025deepsense}.

\subsubsection{Task breadth beyond detection and forecasting}
V2X cooperation also spans identity continuity and cross-view association. DAIR\textnormal{-}V2XReID contributes a real-world vehicle–infrastructure Re-ID dataset built from vehicle and roadside cameras across many intersections, enforcing tight time-matching during acquisition—useful for evaluating how identity evidence survives occlusion and viewpoint change in cooperative settings~\cite{wang2024dair}. Such additions complement perception/forecasting corpora and encourage evidence contracts that keep IDs, timestamps, and camera roles explicit~\cite{wang2024dair}. V2I-focused efforts (e.g., V2I-CARLA and V2I-BEVF) further expand infrastructure-side vantage points and BEV fusion practices~\cite{wang2022v2i,xiang2023v2i}. Together, these resources suggest a unifying design for SPD-era datasets and evaluations:
\begin{itemize}[leftmargin=*]
\item \textbf{Sensing layer (raw $\rightarrow$ tokens):} release synchronized multi-agent streams with per-sample timing, pose, and \emph{continuity} flags; include link/load descriptors or surrogates (e.g., message rate, drop patterns) to emulate PQoS (cf.\ DeepSense-V2V sequencing/metadata)~\cite{morais2025deepsense}.
\item \textbf{Perception layer (predictive structure):} provide trajectories, occupancy/flow, and intent distributions with calibration status and uncertainty summaries, across \emph{cooperative} and ego views (cf.\ V2X\textnormal{-}Seq tasks and vector-map/traffic-light context)~\cite{yu2023v2x}.
\item \textbf{Decision layer (risk/usefulness):} log trigger candidates and outcomes with commitment-window timing and source provenance; evaluate earliness and agreement under link regimes using simulation knobs (cf.\ V2X\textnormal{-}Sim testbed)~\cite{li2022v2x}.
\end{itemize}

\subsection{Embodied and Predictive Intelligence in SPD Loops}
\label{subsec:embodied_spd}

While existing cooperative frameworks excel at information sharing, SPD systems are moving toward \emph{closed-loop, decision-oriented intelligence}, where perception, prediction, and action evolve as co-adaptive processes. Predictive intelligence transforms the SPD loop from passive evidence fusion into proactive behavior reasoning: agents move beyond forecasting what is likely to happen to assessing what \emph{can} and \emph{should} happen under evolving constraints, goals, and social dependencies. This perspective connects physical feasibility, anticipatory reasoning, and cooperative decision-making within a closed-loop reasoning cycle.

\subsubsection{Decision-Oriented Forecasting and Actionable Evidence}
Recent studies emphasize that forecasting in autonomous driving should no longer be treated as an isolated sequence extrapolation stage, but as a provider of \emph{actionable evidence} for downstream decision-making~\cite{wang2024deepaccident,zhou2023interaction}. Instead of outputting trajectories evaluated by geometric error, a growing body of work embeds prediction into end-to-end or planner-aware systems. End-to-end architectures based on imitation or affordance learning map raw sensory inputs to high-level driving indicators or control commands—such as lane-relative position, distance-to-collision, or discrete maneuver choices—that can be directly consumed by low-level controllers~\cite{cui2022coopernaut,liu2024eco}. Likewise, joint perception–prediction frameworks in bird’s-eye view fuse detection, tracking, and motion forecasting into unified networks whose outputs are planner-ready trajectories or occupancy fields~\cite{chang2023bev,ruan2024learning}, narrowing the gap between what is perceived and what the driving policy can act upon.

From an SPD perspective, these decision-oriented designs illustrate how forecasting can be tailored to the requirements of the decision layer. Predicted futures are time-stamped, calibrated, and associated with uncertainty summaries, enabling multiple agents and viewpoints to reason over a common probabilistic description of the scene~\cite{wang2024deepaccident,li2024spatiotemporal}. Forecast representations are further aligned with geometric and semantic constraints, such as drivable areas, signal phases, and right-of-way rules, so that each trajectory hypothesis encodes what an agent is likely to do and what is feasible under current conditions~\cite{chang2024cav}. Several frameworks also optimize downstream objectives, including collision risk, comfort, and rule compliance, allowing predictive outputs to be structured along decision-relevant dimensions and geometric consistency~\cite{zhou2023interaction}.

In SPD-coordinated V2X systems, this trend suggests that the role of motion forecasting is to maintain a stream of structured future hypotheses that can be directly transformed into cooperative risk estimates and intervention candidates~\cite{shi2024mtr,ribeiro2023evaluation}. Rather than serving as a detached pre-processing block, forecasting becomes a decision-oriented service: it aggregates multi-agent evidence from the sensing layer, organizes it into calibrated and constraint-aware scenarios at the perception layer, and exposes these scenarios to the decision layer in a form that is immediately usable for triggering warnings, negotiating maneuvers, or updating shared cost landscapes across vehicles and infrastructure~\cite{chang2023bev,cheng2025driving}.

{Building on decision-oriented forecasting, Foundation-model-driven autonomous driving is also reshaping future SPD loops~\cite{wu2026foundation,wu2024prospective}. Large vision-language models, world models, and end-to-end planning frameworks can integrate sensor observations, maps, traffic rules, and natural-language instructions into unified reasoning modules~\cite{you2026v2x,guan2024world}. Within cooperative safety intelligence, their potential role lies in decision-layer interpretation, scenario-level risk reasoning, and human-understandable planning support~\cite{chen2024end}. However, their deployment in V2X safety systems still requires communication-aware grounding, latency-bounded inference, uncertainty calibration, and verifiable safety constraints.}

\subsubsection{Affordance Reasoning and Causal Prediction}
Accident anticipation studies reveal the need to embed \emph{causal} and \emph{affordance-based} reasoning into perception–decision pipelines. 
Cognitive Traffic Accident Anticipation~\cite{li2024cognitive} introduces a hierarchical reasoning model that links visual semantics, event causality, and temporal dependencies, moving beyond black-box prediction toward interpretable causal inference. 
Dynamic spatio-temporal attention networks~\cite{karim2022dynamic} and visual–temporal surveys~\cite{fang2023vision} further validate that early cues—such as interaction topology and trajectory inflection—encode feasible action spaces critical for risk estimation. 
Within SPD cooperation, such affordance reasoning ensures that perception outputs are not merely geometric observations but structured \emph{evidence of feasibility}. 
Each forecast thereby acts as a causal hypothesis: which maneuvers are physically possible, socially permissible, and temporally plausible. 
In embodied SPD loops, affordance-driven causal prediction becomes the connective tissue linking environmental understanding and intervention readiness.

\subsubsection{Adaptive and Cognitively Grounded Forecasting for SPD}
The LATTE framework~\cite{zhang2025latte} illustrates how embodied prediction can integrate cognitive accessibility and real-time adaptability. By combining lightweight attention with a vision–language alert system, LATTE bridges predictive modeling and human interpretability—delivering verbal hazard explanations that close the loop between machine anticipation and human comprehension. This cognitive grounding aligns with Hussain \textit{et~al.}~\cite{hussain2021autonomous}, who conceptualize autonomous driving as oscillating between \emph{anticipation} (foresight) and \emph{apprehension} (trust and safety perception). 
Within SPD loops, such adaptive intelligence motivates forecasting modules to modulate perception fidelity, communication load, and decision urgency in response to uncertainty and environmental entropy.
From this perspective, embodied prediction functions as a control core for cooperative safety intelligence, coordinating what to sense, when to communicate, and how to act under time- and trust-limited conditions.

\subsection{Trust, Transparency, and Human-in-the-Loop Cooperation}
\label{subsec:trust_hitl}

Beyond technical accuracy, the sustainability of cooperative safety intelligence relies on \emph{trustworthiness, transparency, and interpretability}.
In future SPD ecosystems, cooperation among human operators, AI agents, and regulators requires a shared understanding of system intent, uncertainty, and responsibility.
This subsection discusses how explainability, confidence communication, and human-in-the-loop design foster reliable cooperation across sensing, perception, and decision layers.

\subsubsection{Explainable and Transparent Cooperation}
Transparency is a prerequisite for trust calibration in cooperative automation. 
Explainable AI (XAI) has emerged as a central enabler, offering human-understandable reasoning across the SPD loop. 
Kuznietsov \textit{et~al.}~\cite{kuznietsov2024explainable} establish that explainability underpins safe and trustworthy autonomous driving by linking data integrity, model interpretability, and agency accountability through the \emph{SafeX} framework, which connects interpretable design, monitoring, and validation within safety assurance. 
Similarly, Zhang \textit{et~al.}~\cite{zhang2024critical} demonstrate that interpretability directly shapes public perception and acceptance of autonomous vehicles, with transparent feedback loops mitigating over- or under-trust. 
Zablocki \textit{et~al.}~\cite{zablocki2022explainability} and Dong \textit{et~al.}~\cite{dong2023did} highlight that feature visualization and causal explanation are indispensable to demystifying deep vision models, thus transforming perception outputs into auditable, human-aligned evidence. 
Collectively, these works reposition XAI not as an auxiliary tool but as a cooperative interface between humans and AI—supporting traceability, fairness, and regulatory compliance in SPD reasoning.

\subsubsection{Confidence, Uncertainty, and Cooperative Assurance}
Transparency extends beyond symbolic explanation to include \emph{quantitative confidence communication}. 
Roitberg \textit{et~al.}~\cite{roitberg2022my} empirically show that displaying confidence levels in cooperative automated driving significantly improves driver understanding and perceived reliability, enabling smoother joint control. 
Parallel research on uncertainty modeling and reinforcement learning~\cite{wu2022uncertainty,shao2025prediction} proposes methods to propagate epistemic and aleatoric uncertainty through perception and planning modules, producing trust-calibrated actions. 
Within the SPD framework, such mechanisms serve as \emph{PQoS} indicators, informing both human partners and machine agents of evidence reliability and decision stability. 
At the system level, Lv \textit{et~al.}~\cite{lv2021deep} demonstrate that digital-twin environments can quantify trust and security through deep learning-driven monitoring, offering a scalable substrate for cooperative validation. Together, these studies lay the foundation for confidence-aware SPD cooperation.

\subsubsection{Human-in-the-Loop Reasoning and Ethical Transparency}
Trust is ultimately grounded in shared accountability. 
Zhang \textit{et~al.}~\cite{zhang2024public} reveal that societal acceptance of autonomous systems depends on how responsibility is perceived under uncertainty—whether errors stem from human oversight, AI decision logic, or cooperative interaction mismatches. 
Integrating human feedback into SPD loops ensures that humans remain active arbiters of intent interpretation and ethical oversight. 
Cognitive approaches such as Cog-TAA~\cite{li2024cognitive} exemplify how driver attention and textual reasoning can be fused with AI perception to enhance both explainability and mutual understanding, bridging machine predictions with human situational awareness. 
Future SPD architectures will benefit from adaptive HITL protocols that adjust explanation granularity, communication bandwidth, and intervention thresholds according to contextual uncertainty and operator state. 
In doing so, the SPD framework evolves toward \emph{reciprocal transparency}: systems that not only act intelligibly but can be interrogated, trusted, and guided by human partners.

\section{Conclusion}
This survey has examined the evolution of V2X-enabled traffic safety through the lens of a unified SPD framework. By articulating how safety emerges from the coupling of distributed sensing, cooperative perception, and coordinated decision-making, the SPD perspective  provides a coherent structure for analyzing the increasingly complex landscape of cooperative safety intelligence.
{The structured corpus construction and SPD-aligned literature mapping provide a consistent basis for organizing the field from sensor-centric studies toward more integrated, cross-layer designs that incorporate multi-modal perception, predictive reasoning, and collaborative intervention.}

Across the literature, three constraints (timing, trust, and human factors) consistently shape whether cooperative insights translate into reliable safety gains, yet persistent gaps remain in alert earliness, uncertainty calibration, inter-agent coordination, and user-centered outcomes. Closing these gaps calls for benchmarks and methodologies that respect realistic timing budgets, communication conditions, and human--machine interactions, while coupling geometric accuracy with safety-critical metrics. Looking forward, mature cooperative safety intelligence will be enabled by SPD-driven fusion that integrates perception and prediction with calibrated uncertainty, PQoS-aware communication strategies that prioritize safety relevance, human-centered intervention design, and standardized datasets and digital twins that support reproducible closed-loop evaluation, transforming V2X cooperation into a dependable foundation for next-generation transportation systems and, ultimately, Zero-Accident Mobility.

\section*{Acknowledgments}
This work was supported by the Science and Technology Development Fund of Macau SAR [0007/2025/RIC, 0122/2024/RIB2, 0215/2024/AGJ, 0074/2025/AMJ, 001/2024/SKL, 0028/2025/AFJ], the Research Services and Knowledge Transfer Office, University of Macau [SRG2023-00037-IOTSC, MYRG-GRG2024-00284-IOTSC], the Shenzhen-Hong Kong-Macau Science and Technology Program Category C [SGDX20230821095159012], the Science and Technology Planning Project of Guangdong [2025A0505010016], National Natural Science Foundation of China [52572354], the State Key Lab of Intelligent Transportation System [2024-B001], and the Jiangsu Provincial Science and Technology Program [BZ2024055].

\bibliographystyle{IEEEtran}
\bibliography{IEEEabrv,main}

\section{Biography Section}

\vspace{-9mm}
\begin{IEEEbiography}
[{\includegraphics[width=1in,height=1.25in,clip,keepaspectratio]{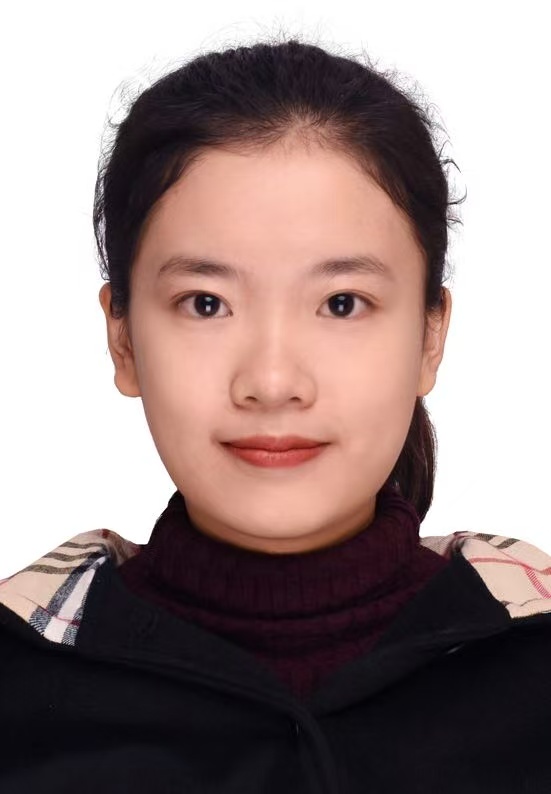}}]
{Jiaxun Zhang} is currently pursuing the Ph.D. degree with the State Key Laboratory of Internet of Things for Smart City and the Department of Civil and Environmental Engineering, Faculty of Engineering, University of Macau. She received the M.S. degree in Integrated Sustainable Design from the National University of Singapore in 2023 and the B.E. degree in Traffic Engineering from South China University of Technology in 2022. Her research primarily focuses on the integration of artificial intelligence with autonomous driving technologies and intelligent transportation systems.
\end{IEEEbiography}

\vspace{-9mm}
\begin{IEEEbiography}[{\includegraphics[width=1in,height=1.25in,clip,keepaspectratio]{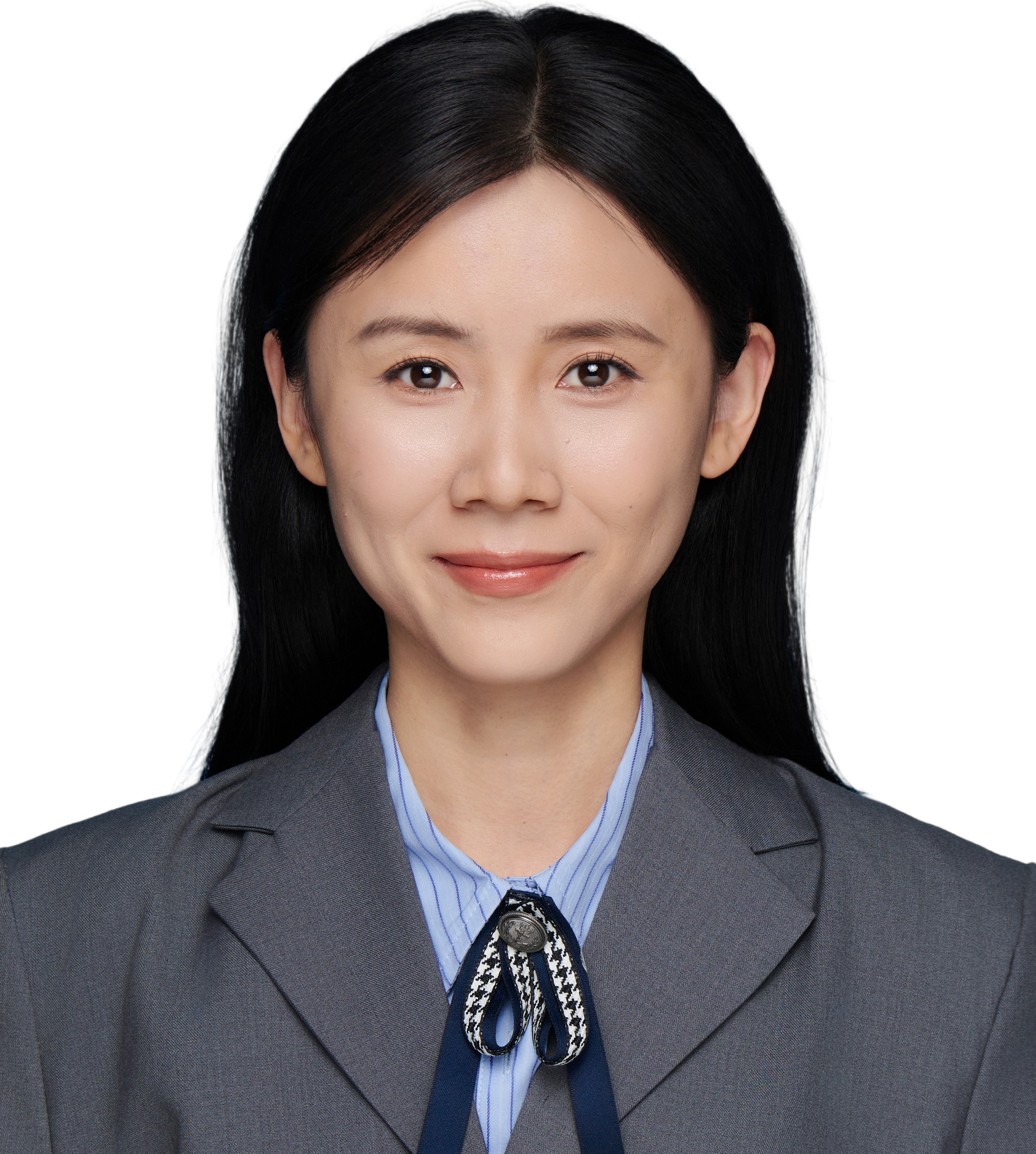}}]{Qian Xu} is currently a Post-Doctoral Fellow with the State Key Laboratory of Internet of Things for Smart City, University of Macau. She holds the Ph.D. degree in Transportation Engineering from Tongji University (2025), the Master's degree in Traffic Information Engineering and Control from Lanzhou Jiaotong University (2021), and the bachelor's degree in Railway Traffic Signal and Control from Lanzhou Jiaotong University (2018). Her research focuses on safety, security and AI techniques for autonomous driving, V2X.
\end{IEEEbiography}

\vspace{-9mm}
\begin{IEEEbiography}
[{\includegraphics[width=1in,height=1.25in,clip,keepaspectratio]{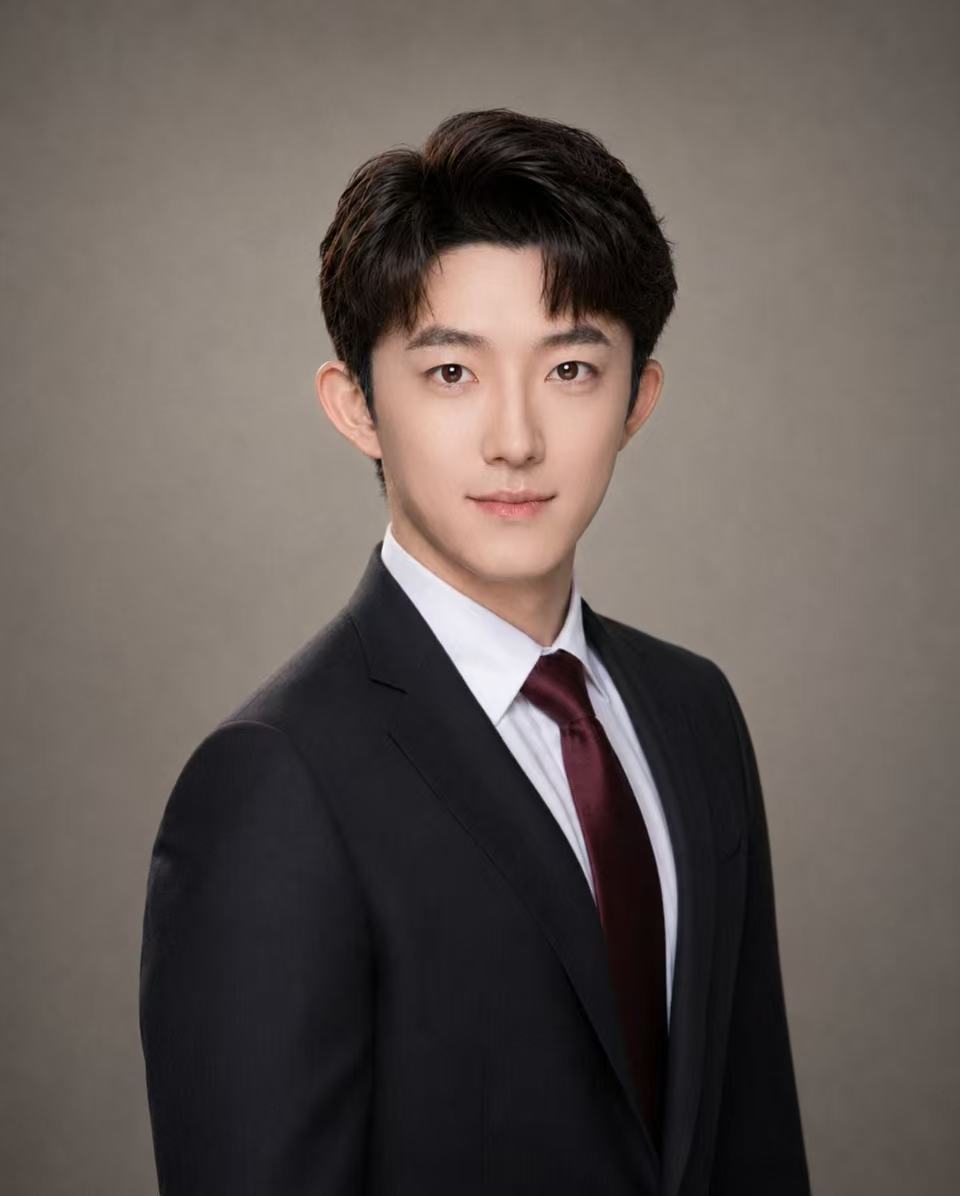}}] {Zhenning Li} (Member, IEEE) received his Ph.D. in Civil Engineering from the University of Hawaii at Manoa, Honolulu, Hawaii, USA, in 2019. Currently, he holds the position of Assistant Professor at the State Key Laboratory of Internet of Things for Smart City, as well as the Department of Civil and Environmental Engineering, Faculty of Engineering, and the Department of Artificial Intelligence, Faculty of Information Science and Computing at the University of Macau. Over his academic career, he has published over 80 papers. His main areas of research focus on the intersection of connected autonomous vehicles and Big Data applications in urban transportation systems. He has been honored with several awards, including the TRB best young researcher award and the CICTP best paper award.
\end{IEEEbiography}

\vspace{-9mm}
\begin{IEEEbiography}
[{\includegraphics[width=1in,height=1.25in,clip,keepaspectratio]{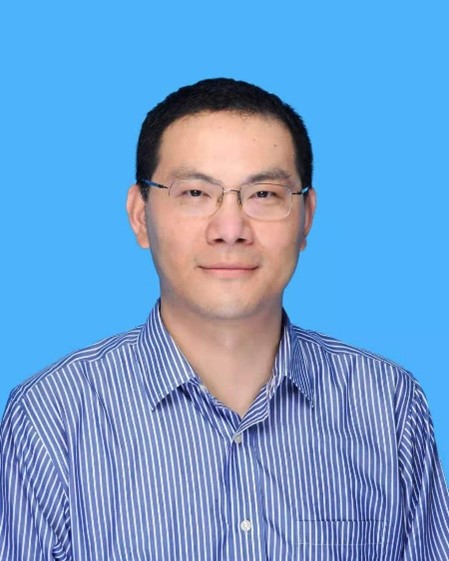}}]{Yuan Wu} (Senior Member, IEEE) is currently a Full Professor with the State Key Laboratory of Internet of Things for Smart City and the Department of Electronic and Communication Engineering, Faculty of Information Science and Computing, University of Macau, Macau SAR, China. His research interests focus on resource management for wireless communications and networks, mobile edge computing and edge intelligence, and integrated sensing and communications. Dr. Wu was a recipient of the Best Paper Award from the IEEE ICC 2016, IEEE TCGCC 2017, IWCMC 2021, and IEEE WCNC 2023. He serves on the Editorial Board of IEEE Transactions on Wireless Communications, IEEE Transactions on Vehicular Technology, and IEEE Transactions on Network Science and Engineering. He is a Distinguished Lecturer of the IEEE Vehicular Technology Society (VTS).
\end{IEEEbiography}

\vspace{-9mm}
\begin{IEEEbiography}
[{\includegraphics[width=1in,height=1.25in,clip,keepaspectratio]{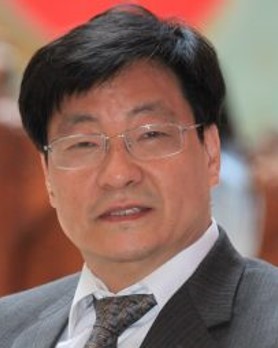}}]{Chengzhong Xu} (Fellow, IEEE) received the Ph.D. degree from The University of Hong Kong in 1993. He is currently a Chair Professor of Computer Science, the Dean of the Faculty of Information Science and Computing, and the Interim Dean of the Faculty of Engineering at the University of Macau. Prior to this, he was with the faculty at Wayne State University, USA, and the Shenzhen Institutes of Advanced Technology, Chinese Academy of Sciences, China. He has published more than 400 papers and more than 100 patents. His research interests include cloud computing and data-driven intelligent applications. He was a recipient or nominee of the Best Paper Awards at ICPP 2005, HPCA 2013, HPDC 2013, Cluster 2015, GPC 2018, UIC 2018, and AIMS 2019. He also received the Best Paper Award at SoCC 2021. He served as the Chair of the IEEE Technical Committee on Distributed Processing from 2015 to 2019.
\end{IEEEbiography}

\vspace{-9mm}
\begin{IEEEbiography}[{\includegraphics[width=1in,height=1.25in,clip,keepaspectratio]{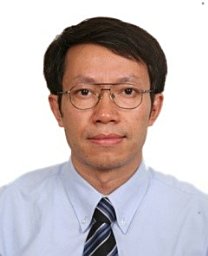}}]{Keqiang Li} received the B.Tech. degree from Tsinghua University, Beijing, China, in 1985, and the M.S. and Ph.D. degrees in mechanical engineering from Chongqing University, Chongqing, China, in 1988 and 1995, respectively. He is currently a Professor with the School of Vehicle and Mobility, Tsinghua University. His main research areas include automotive control systems, driver assistance systems, and networked dynamics and control. He is leading the national key project on Intelligent and Connected Vehicles (ICVs) in China. He has authored more than 200 papers and is a co-inventor of over 80 patents in China and Japan. He has served as a Member of the Chinese Academy of Engineering, a Fellow of the Society of Automotive Engineers of China, a member of the editorial board of the \textit{International Journal of Vehicle Autonomous Systems}, Chairperson of the Expert Committee of the China Industrial Technology Innovation Strategic Alliance for ICVs (CAICV), and Chief Technology Officer (CTO) of the China ICV Research Institute Company Ltd. (CICV). He has been a recipient of the Changjiang Scholar Program Professorship, the National Award for Technological Invention in China, among others.
\end{IEEEbiography}

\end{document}